\documentclass[journal]{IEEEtran}
\usepackage{stfloats}

\usepackage{enumerate}
\usepackage{algorithm,algpseudocode}
\usepackage{setspace,amsmath,latexsym,cite,amssymb,epsfig,amsfonts}
\usepackage{url,cite}
\usepackage{graphicx}
\usepackage{psfrag}
\usepackage{footmisc}
\usepackage{multirow}
\usepackage{color}
\usepackage{multicol}
\usepackage{mathtools}
\usepackage{subcaption}
\usepackage{graphicx}
\usepackage{amssymb}
\usepackage{amsmath}
\usepackage{epstopdf}
\usepackage{geometry}
\usepackage{float}
\usepackage{balance}
\usepackage{makecell}
\usepackage{changepage}

\algnewcommand\algorithmicforeach{\textbf{for}}
\algdef{S}[FOR]{For}[1]{\algorithmicforeach\ #1\ \algorithmicdo}

\twocolumn

\makeatother

        \makeatletter
        \def\fps@eqnfloat{!t}
        \def\ftype@eqnfloat{4}
        
        \newenvironment{eqnfloat*}
               {\@dblfloat{eqnfloat}}
               {\end@dblfloat}
        \makeatother
\allowdisplaybreaks[4]

\begin{document}

\title{Deep Learning-Based Traffic-Aware Base Station Sleep Mode and Cell Zooming Strategy in RIS-Aided Multi-Cell Networks\\
}

\author{Shuo Sun, \IEEEmembership{Student Member, IEEE}, Chong Huang, \IEEEmembership{Member, IEEE}, Gaojie Chen, \IEEEmembership{Senior Member, IEEE}, \\ Pei Xiao, \IEEEmembership{Senior Member, IEEE}, and Rahim Tafazolli, \IEEEmembership{Fellow, IEEE}

\thanks{This work was supported in part by the U.K. Engineering and Physical Sciences Research Council under Grant EP/X013162/1, and in part by the Fundamental Research Funds for the Central Universities, Sun Yat-sen University, under Grant No.24hytd010.}
\thanks{S. Sun is with 5GIC \& 6GIC, Institute for Communication Systems (ICS), University of Surrey, Guildford, GU2 7XH, United Kingdom, and also with National Innovation Center for Digital Fishery, China Agricultural University, Beijing 10083, China. Email: s.sun@surrey.ac.uk.}
\thanks{C. Huang, P. Xiao and R. Tafazolli are with 5GIC \& 6GIC, Institute for Communication Systems (ICS), University of Surrey, Guildford, GU2 7XH, United Kingdom. Email: \{chong.huang, p.xiao, r.tafazolli\}@surrey.ac.uk. (Corresponding author: C. Huang)}
\thanks{G. Chen is with the School of Flexible Electronics (SoFE) \& State Key Laboratory of Optoelectronic Materials and Technologies (OEMT), Sun Yat-sen University, Shenzhen, Guangdong 518107, China. Email: gaojie.chen@ieee.org.}
}

\maketitle

\begin{abstract}
Advances in wireless technology have significantly increased the number of wireless connections, leading to higher energy consumption in networks. Among these, base stations (BSs) in radio access networks (RANs) account for over half of the total energy usage. To address this, we propose a multi-cell sleep strategy combined with adaptive cell zooming, user association, and reconfigurable intelligent surface (RIS) to minimize BS energy consumption. This approach allows BSs to enter sleep during low traffic, while adaptive cell zooming and user association dynamically adjust coverage to balance traffic load and enhance data rates through RIS, minimizing the number of active BSs. However, it is important to note that the proposed method may achieve energy-savings at the cost of increased delay, requiring a trade-off between these two factors. Moreover, minimizing BS energy consumption under the delay constraint is a complicated non-convex problem. To address this issue, we model the RIS-aided multi-cell network as a Markov decision process (MDP) and use the proximal policy optimization (PPO) algorithm to optimize sleep mode (SM), cell zooming, and user association. Besides, we utilize a double cascade correlation network (DCCN) algorithm to optimize the RIS reflection coefficients. Simulation results demonstrate that PPO balances energy-savings and delay, while DCCN-optimized RIS enhances BS energy-savings. Compared to systems optimised by the benchmark DQN algorithm, energy consumption is reduced by 49.61\%.
\end{abstract}


\section{Introduction}
In recent years, with the widespread popularity of wireless terminal devices such as smart vehicles and a large number of IoT devices \cite{10049164,9592698}, the demand for wireless communication networks has increased significantly. According to global system for mobile communications (GSMA) statistics, the number of active connections established through smartphones and feature phones in Europe has exceeded 900 million. By 2030, the ownership rate of such devices will reach 91$\%$ \cite{gsma2023mobile}. The high demand has led to increased energy consumption and large-scale emissions of greenhouse gases, exacerbating
global warming \cite{shinkuma2021smarter}. To address this issue, the concept of green communication has been introduced to reduce energy consumption in the telecommunications industry. Base stations (BSs), which are vital infrastructures in mobile wireless networks, account for more than 80\% of the energy in mobile cellular networks \cite{10285118}. Therefore, research on energy-saving technologies for BSs is crucial and imperative.

BS sleep mode (SM) is one of the technologies that can significantly reduce the energy consumption of mobile cellular networks without altering the current network structure, resulting in substantial cost savings \cite{9729881}. It operates at the sub-component level, deactivating certain sub-components when the BS experiences low traffic loads, effectively putting the BS into a sleep state. On the other hand, compared with other energy-saving technologies such as the application of renewable energy, SM can be implemented based on existing network infrastructure, thus it is considered a cost-effective energy-saving technology \cite{wang2023base}. The design of SM involves various factors, such as the characteristics of dynamic traffic loads. For instance, the traffic variation of the tidal pattern was considered in \cite{lin2021data}, which proposed a bidirectional long short-term memory (BLSTM) network to predict future traffic for BS switching on/off strategy. The authors of \cite{piovesan2020joint} studied the impact of traffic changes on SM strategies and combined them with renewable power to decide whether to switch the BS on or off. In \cite{8122299}, the authors investigated the collaborative state management of multiple BSs and used the spatio-temporal characteristics of traffic to design a BS joint state management and clustering algorithm based on the arrival traffic queue. Furthermore, energy efficiency (EE) and delay are key factors considered in the development of SM. Three schemes were considered in \cite{9109301}: the isolated scheme, the cooperative scheme, and the hybrid scheme, to trade off energy consumption and mean delay. Similarly, SM was considered in \cite{8462774} to minimize BS energy consumption. The above method effectively achieves BS energy-savings. However, these works only consider
the simple switching on/off strategy of BSs, which may not fully exploit the energy-saving potential of SM.

Additionally, multi-level SM has been proposed in \cite{salem2017advanced} which measured the energy consumption of BS components and proposed four SM levels, which are known as advanced sleep modes (ASMs): extra deep sleep, deep sleep, light sleep, and micro-sleep. The energy consumption increases and the reactivation time decreases as the state transitions from extra deep sleep to micro-sleep. The studies in \cite{chang2019energy,israr2023renewable,razzac2023advanced,renga2023trading} have utilized this approach to further investigate energy-savings for BSs. The study in \cite{chang2019energy} associated four SM modes of ASM with user demands. If the demand is delay-sensitive and rate-sensitive, the BS will switch to light sleep; if not, it will switch to deep sleep. The authors of \cite{israr2023renewable} proposed a traffic-aware offloading approach with ASM that was considered to match the dynamics of available energy changes with the arrival of service requests. Multi-BS ASM was proposed in \cite{razzac2023advanced}, which trades off delay while achieving energy-savings, and considered whether the interference between multiple BSs affects the wake-up of sleeping BSs. Similarly, trade-offs of ASM and delay were studied in \cite{renga2023trading}.

However, existing research mainly focused on the trade-off between energy and delay in ASM, especially in deciding when ASM enters which sleep level and the duration of each level, while less attention has been paid to integrating other energy-saving technologies to further reduce energy consumption. To address this issue, we begin to think about the coverage of BSs. Maintaining fixed coverage all the time will result in wasted resources and degraded performance due to over-coverage in high-traffic load areas and under-coverage in low-load areas. To solve this problem, an innovative energy-saving strategy was proposed by the authors \cite{5621970} which is known as ``cell zooming''. This concept involves dynamically adjusting the size of each cell in response to traffic fluctuations, facilitating a cooperative framework among multiple BSs to enhance network efficiency and reduce delay.  There are two levels of cell zooming: zooming out and zooming in \cite{10042987}. For instance, in a scenario where a large number of users around the BS causes significant traffic load and subsequent congestion, the BS will zoom in to reduce cell size, thereby serving fewer users. Adjacent BSs with low traffic load will zoom out to take over the offloaded users, achieving load balancing. The studies in \cite{8025626, khaled2020green, han2022energy} used cell zooming to achieve energy-savings for BSs. The authors of \cite{8025626} proposed the adaptive cell zooming scheme and defined the cell zooming factor to express the cell coverage. A spectrum-based cell zooming was combined with software radio technology in \cite{khaled2020green} to design a green traffic framework to minimize grid energy consumption and maximize throughput. The studies in \cite{han2022energy} used cell zooming to achieve energy-efficient deployment in mobile edge cloud (MEC) networks.

Furthermore, the reconfigurable intelligent surface (RIS) is proposed as a revolutionary technology that utilizes a large number of low-cost passive elements capable of adjusting phase-shifted reflected signals to achieve energy-savings. This technology could make data transmission rates faster by enhancing the signal and providing more opportunities for the BS to enter the sleep state, which is another way to achieve a higher degree of energy-savings \cite{8910627}. The studies in \cite{10103550,ren2022energy,yaswanth2023energy,9352948} employed RIS-assisted methodologies to enhance the signal, thereby facilitating a reduction in energy consumption. The authors of \cite{10103550} proposed that RIS is a potential technology that can improve the energy efficiency of the transmitter by reflecting and strengthening the signal. The authors of \cite{ren2022energy} adjusted the phase offset of the RIS reflection unit to enhance the passive beamforming of the signal, thereby significantly improving the energy transfer efficiency and reducing the total system energy consumption. Similarly, the combination of RIS with simultaneous wireless information and energy transmission was proposed in \cite{yaswanth2023energy}, which optimized the active beamforming matrix and the passive beamforming matrix of RIS to achieve significant energy-savings. The studies in \cite{9352948} proposed using RIS to enhance cell-free MIMO system energy efficiency and maximize it through hybrid beamforming. However, the performance of RIS-assisted communications has not been considered for enhancing EE in green BS communication scenarios.

Building upon the aforementioned technology, SM, cell zooming, user association, and RIS techniques facilitate energy conservation for future sixth-generation (6G) green communications. Nonetheless, optimizing energy-efficient communication with BSs utilizing SM, supported by technologies such as cell zooming, user association, and RIS is a complex non-convex problem. Recently, deep reinforcement learning (DRL) has demonstrated significant potential in the field of energy conservation in wireless communications \cite{10104154,9715791,9340607,9295428,8735834}. The authors of \cite{10104154} proposed a method to optimize the power management of multiple BSs using deep Q-learning (DQN) to improve the long-term network energy efficiency. The authors of \cite{9715791} utilized DQN to determine the appropriate SM level of BS to achieve the lowest energy consumption. A traffic-aware SM framework was considered in \cite{9340607} to model the network environment as a Markov decision process (MDP) which was solved using the deep deterministic policy gradient (DDPG). However, due to the design of DDPG, it is mainly applicable to continuous action spaces and is limited when dealing with discrete or mixed action spaces (including both discrete and continuous action spaces). In this regard, Asynchronous Advantage Actor-Critic (A3C) provides a more suitable optimization method, especially in these more complex environments with discrete action spaces \cite{9366889}. The studies in \cite{9868213} utilised A3C to obtain the location deployment of all UAVs and BSs to improve energy efficiency. Additionally, compared to A3C and DDPG, Proximal Policy Optimization (PPO) offers a more stable and easier-to-tune alternative \cite{10463053}. PPO introduces a clipping mechanism to prevent overly large policy updates and allows for sample reuse, thereby improving sample efficiency during training. Moreover, due to the short-term nature of RIS optimization and the inability to acquire all possible channel coefficients, the double deep learning network model, proposed in \cite{9115725}, offers a promising solution that focuses on optimizing the RIS by dynamically adjusting the reflection coefficients based on real-time channel conditions.

In this paper, we explore an optimization approach for energy conservation in wireless communication systems. Although the standard ASM has exhibited commendable energy-saving performance in a lot of studies, it falls short of addressing the intricacies of multi-cell cooperation. To tackle this issue, we introduce cell zooming to adaptively adjust the cell size. To be specific, cell zooming and user association allow the cell to adaptively adjust its size to coordinate with adjacent BSs to maximize the number of BSs in sleep states through zooming in and zooming out. Furthermore, to further achieve energy-savings, we introduced the RIS-assisted method to speed up the data transmission rate by reconstructing the wireless environment between active BSs and users. This framework enables BSs to maximize their sleep duration thereby minimizing the energy consumption of BSs. Thus, we propose a sleep mode approach supported by cell zooming, RIS, and user association to minimize energy consumption in wireless communication systems. Moreover, we model the multi-cell wireless environment as an MDP and employ the pre-trained proximal policy optimization (PPO) algorithm and double cascade correlation network (DCCN) for the proposed complicated non-convex optimization problem. The main contributions of this work are summarized as follows:
\begin{itemize}
  \item We first propose the integration of collaborative SM technology, adaptive cell zooming, user association and RIS in multi-cell networks. BSs can enter low-power sleep states during low traffic load periods to reduce energy consumption. The adaptive cell zooming and user association mechanism ensures that the active BSs can expand their coverage areas to seamlessly take over the service regions of the sleeping BSs. When the traffic load level is high, the coverage size can be reduced to balance traffic with adjacent BSs that have low traffic load levels. Furthermore, RIS accelerates data transmission by enhancing the signal link under the same transmit power, providing more opportunities for BSs to enter SM.
  \item To effectively manage the proposed network, we first model the problem as an MDP. Minimizing the overall energy consumption in the multi-cell network while ensuring the delay requirement is a complicated non-convex problem. Thus, the PPO algorithm is employed to address this issue. Furthermore, we employ the DCCN to optimize RIS reflection coefficients in real-time based on the current channel state information, ensuring that the system adapts efficiently to varying traffic loads and channel conditions.
  \item We comprehensively evaluate the proposed energy-saving scheme through simulations, comparing it with benchmarks. Results demonstrate that the proposed energy-saving scheme, which uses PPO optimization and DCCN optimization, has the best energy-saving performance under delay constraint. 

\end{itemize}

The remainder of this paper is summarized as follows. Section \ref{systemmodel} explains the system model and the problem formulation. The creation of the MDP, the PPO-based algorithm, and the DCCN-based algorithms are introduced in Section \ref{sectionalgori}. Section \ref{sectionresult} analyzes the performance of the energy-saving system and the proposed algorithms. Finally, Section \ref{secconclu} concludes this paper.

\begin{figure*}[t!]
\centering
\includegraphics[width=14cm]{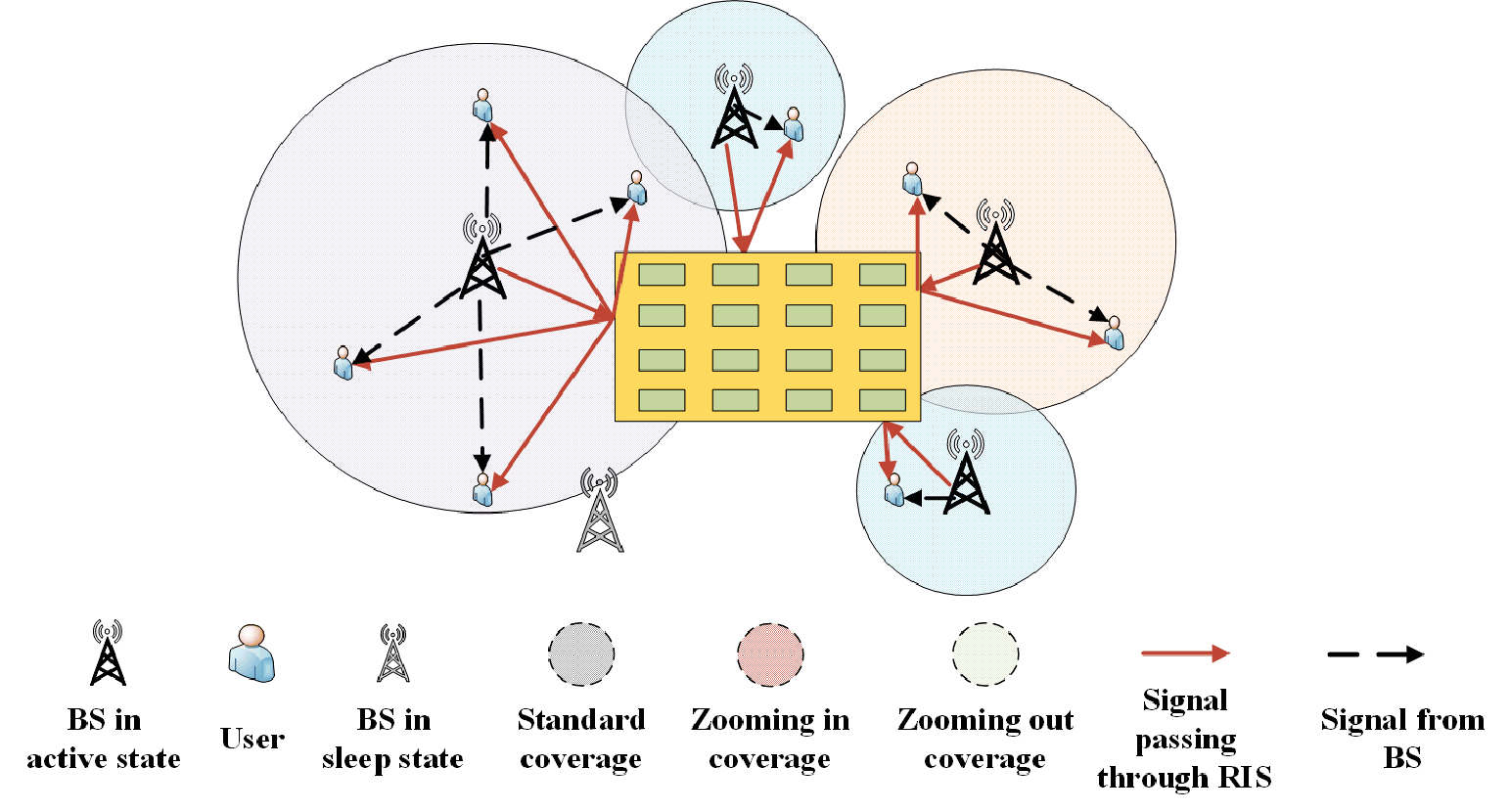}
\caption{System model of an RIS-assisted multi-cell network with SM, cell zooming and user association.}  
\label{systemodel}  
\end{figure*}

\section{System Model}
\label{systemmodel}

\begin{table}[!t]
	\small
	\vspace{5pt}
	\caption{Key notations.}
	\label{notations}       
	\newcommand{\tabincell}[2]{\begin{tabular}{@{}#1@{}}#2\end{tabular}}
	\centering
	\begin{tabular}{|c|c|}
		\hline Notations & Definitions  \\
            \hline $M$ & Number of BSs \\
		\hline $N$ & Number of users\\
		\hline $S_m$ & SM state for ${K}_{m}$     \\
		\hline $C_m$ & Cell zooming level of each BS\\
		\hline $\mathcal{Z}_{M,N}$ & \makecell{User association indicator}\\ \hline
        ${h}_{g,m}$ &\makecell{Channel coefficient between the antenna \\ of $m$-th BS ${K}_{m}$ and the $g$-th RIS element} \\ \hline
        ${h}_{n,g}$ &\makecell{Channel coefficient between the $g$-th \\ RIS element and antenna of $n$-th user ${V}_{n}$ } \\ \hline
        ${h}_{m,n}$ &\makecell{Channel coefficient between BS $K_m$ and user ${V}_{n}$ } \\ \hline
        ${\alpha_1}$ &\makecell{Path loss exponent for LOS } \\ \hline
        ${\alpha_2}$ &\makecell{Path loss exponent for NLOS} \\ \hline
        ${\alpha_3}$ &\makecell{Path loss exponent for Rayleigh fading} \\ \hline

        $\mathbf{\Theta}$ & Phase shift matrix of RIS \\ \hline ${P}_{m}$ & Power of ${K}_{m}$\\
		\hline
         ${R}_{m,n}$ & Transmission rate between ${K}_{m}$ and ${V}_{n}$ \\
        \hline $\mathcal{D}_{i}(t)$ & Transmission delay \\
	\hline ${E}_{\tau}(t)$ & Energy consumption in a
        time slot \\
        \hline ${E}_{K_m}(t)$ & Energy consumption for $K_m$ \\ \hline
        $E_{\text{system}}(t)$ & Total energy consumption of the system\\ \hline
        $\mathcal{U}$ & State space of MDP\\ \hline
        $\mathcal{A}$ & Action space of MDP\\ \hline
        $R_{s, a}(t)$ & Reward function \\ \hline
	\end{tabular}
\end{table}

We consider an RIS-assisted multi-cell network consisting of $M$ BSs and $N$ users, each equipped with a single antenna \footnote{Considering that a single antenna can achieve the lowest power consumption for BSs while satisfying rate requirements \cite{6514953}, and since antenna selection and multiple-input and multiple-output (MIMO) are not the focus of this work, the trade-offs between multi-antenna setups and MIMO will be addressed in future works.}, and an RIS with $G$ elements, as shown in Fig. \ref{systemodel}. The BS set is defined as \( \left\{{K}_{1}, {K}_{2}, {K}_{3}, \dots, {K}_{M} \right\} \), and the $m$-th BS can be represented by $K_m$. The user set is defined as $\left \{{V}_{1},{V}_{2},{V}_{3}, \dots,{V}_{N} \right \}$, and the $n$-th user can be represented by $V_n$.

\subsection{Base Station Model}

\begin{figure}[t!]
\centering
\includegraphics[width = \columnwidth]{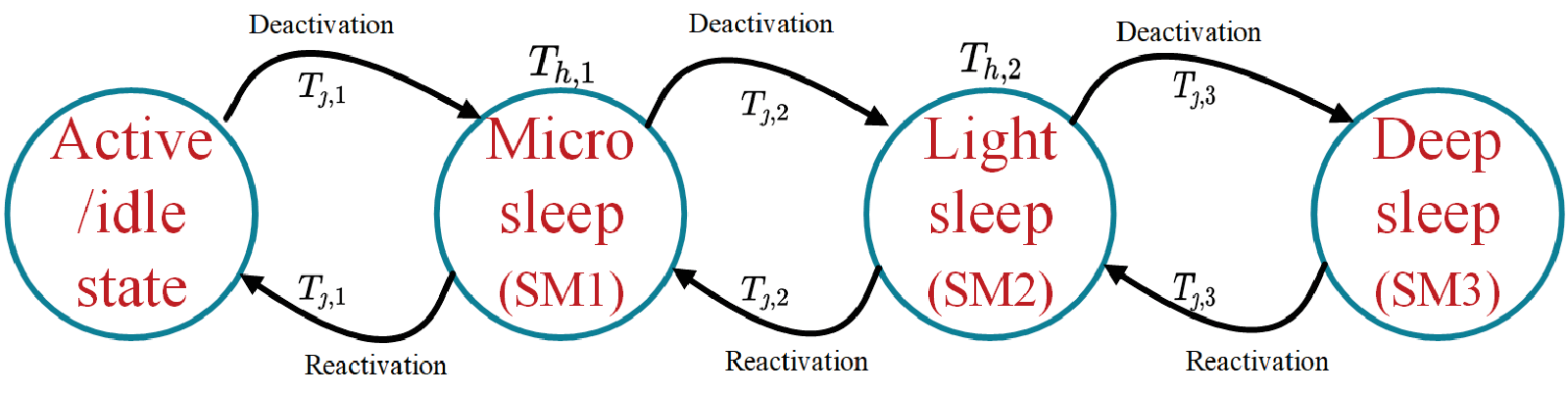}
\caption{Transitions between five states of SM.}
\label{trans}
\end{figure}

ASM is usually divided into four levels, which are created at the level of hardware sub-components and determined by their similar reactivation/deactivation times. Since our optimization goal is to trade off energy consumption with delay, the extra deep sleep in ASM is not considered due to its high delay \cite{masoudi2022digital}. In this paper, the SM states for all BSs are considered to consist of $L$ states, with $L$ being 5, these states include: active state (AS), idle state (IS), micro sleep state (SM1), light sleep state (SM2), and deep sleep state (SM3), which can be represented by $S_m \in \mathcal{S}_M = \{S_1, S_2, ..., S_M\}$, where $S_m = \left \{s_1, s_2, \dots, s_5 \right \}$ denotes the SM states for the BS $K_m$, $m \in \{1, 2, ..., M\}$, $s_1$ is the active state, $s_2$ is the idle state, $s_3$ is micro sleep, $s_4$ is light sleep, and $s_5$ is deep sleep. Fig. \ref{trans} shows the transition between the five states. Different SM states have their own fixed deactivation and reactivation times, and transitions between SM states must follow a defined sequence. This sequence acts as a buffer, allowing the BS to adapt and prepare for entering a deeper sleep state. Additionally, there is a hold time in both the micro sleep and light sleep states.
As shown in Fig. \ref{systemodel}, there is a BS that is not circled, which indicates that the BS is in a sleep or idle state.

\begin{figure}[t!]
\centering
\includegraphics[width = \columnwidth]{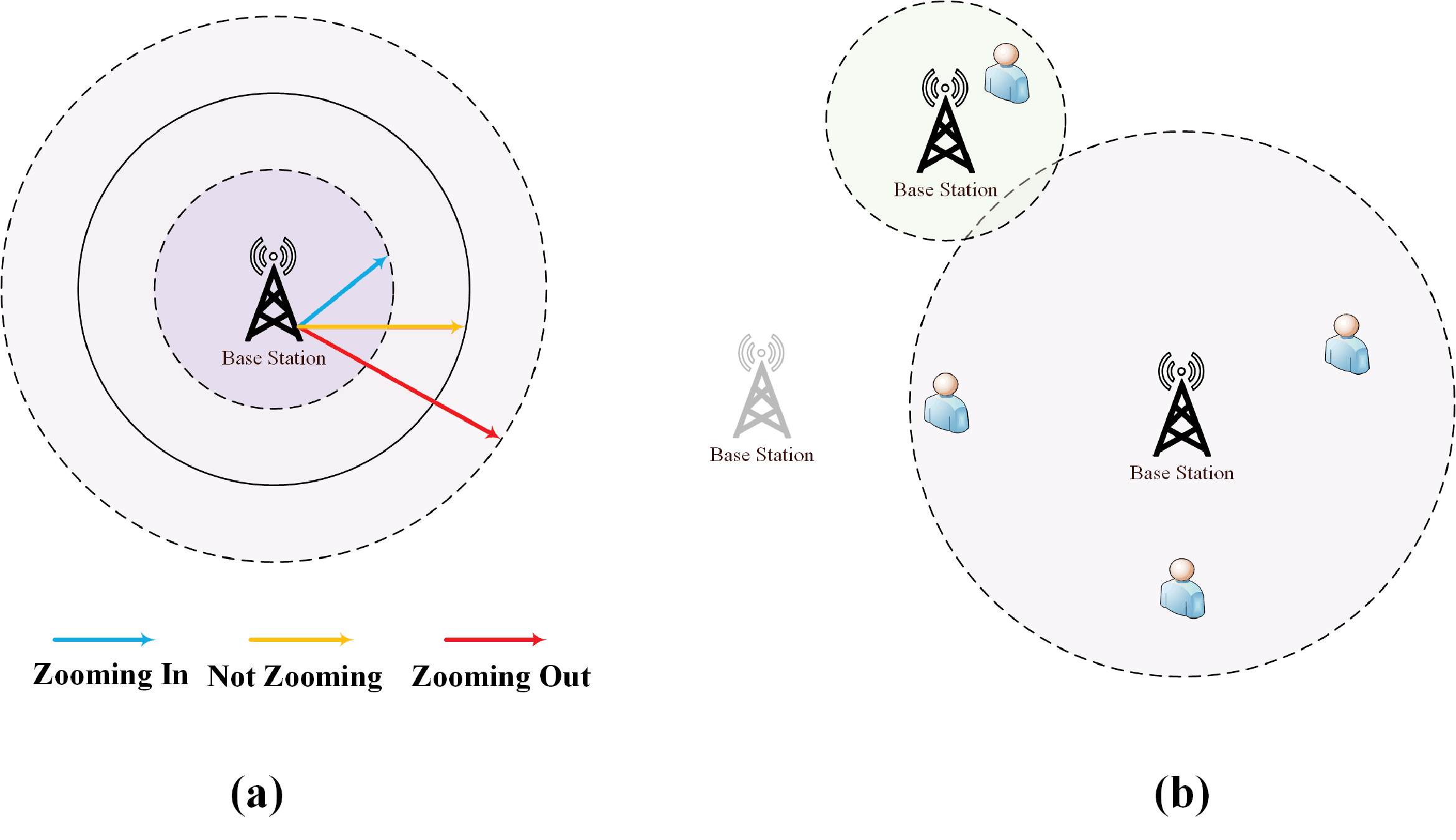}
\caption{Cell zooming and user association for multi-cell cooperation.}
\label{cellzooming}
\end{figure}

For the cell zooming of BSs, there are three levels: zooming in, not zooming, and zooming out, which can be denoted as $C_m \in \mathcal{C}_M = \{C_1, C_2, ..., C_M\}$, where $C_m = \left \{ c_0, c_1, c_2\right \}$ denotes the cell zooming level for the BS $K_m$, $m \in \{1, 2, ..., M\}$, $c_0$ is zooming in, $c_1$ is not zooming, and $c_2$ is zooming out. Fig. \ref{cellzooming}(a) illustrates cell zooming for a BS and Fig. \ref{cellzooming}(b) illustrates user association in cell zooming. There are three circles with different sizes in Fig. \ref{cellzooming}(a), which represent the different cell zooming levels, the largest coverage circle means zooming out, the medium circle means not zooming, and the smallest circle means zooming in. Furthermore, the BS will only perform cell zooming when it is in an active state. This is because when ${K}_{m}$ is not in an active state, it does not have coverage.

User association is reflected in the cooperation of multiple cells as shown in Fig. \ref{cellzooming}(b). There are two user assistance scenarios: when the traffic load of ${K}_{m}$ at a low level and adjacent BSs have a high-level traffic load, the high-level traffic load of the adjacent BSs may cause traffic congestion and increase the delay. To avoid this, the user association mechanism and cell zooming can achieve load balancing that allows adjacent BSs to zoom in to reduce part of their own traffic loads, while ${K}_{m}$ zooming out to take over these traffic loads, thereby reducing the traffic load pressure of adjacent BSs. Conversely, when ${K}_{m}$ has a high-level traffic load and adjacent BSs have a low-level traffic load, ${K}_{m}$ will zoom in while adjacent BSs zoom out to balance the load, ensuring quick task completion and reduced delay. A user can only be served by one BS, and a BS can serve multiple users. The association between BSs and users is denoted by $\mathcal{Z}_{M,N} \in \{Z_{1,1}, Z_{1,2}, ..., Z_{M,N}\}$, where $Z_{m,n} \in \{0,1\} $, $m \in \{1, 2, ..., M\}$, $n \in \{1, 2, ..., N\}$, when $Z_{m,n} = 1 $ indicates that user $V_{n} $ is served by BS $K_{m} $, and $Z_{m,n} = 0 $ indicates that user $V_{n} $ is not served by BS $ K_{m} $. Therefore, user association can be optimized by adjusting $Z_{m,n}$.

\subsection{Transmission and Power Mode}
\begin{figure}[t!]
\centering
\includegraphics[width = 0.7\columnwidth]{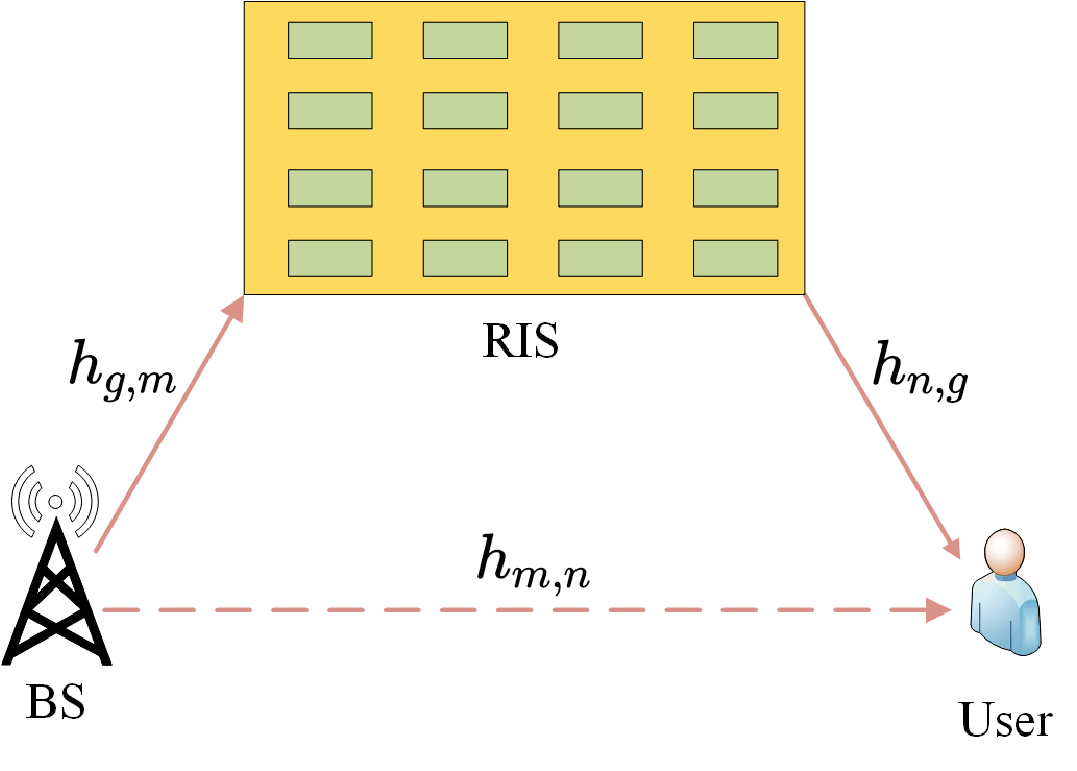}
\caption{RIS-assisted communication scenarios.}
\label{RIS-ass}
\end{figure}

RIS is an emerging energy-saving technology, which consists of a large number of low-cost passive reflective units. Each unit can be independently controlled to adjust the amplitude and phase shift of the incident signal to reflect the signal to the receiving end, thereby reconstructing the wireless propagation environment \cite{9947328}. Fig. \ref{RIS-ass} shows the RIS-assisted communication scenarios. We assume that RIS has $G$ reflecting elements and adjacent cells are not far from each other, so it is assumed that they share an RIS. The channel coefficient between the antenna of the $m$-th BS ${K}_{m}$ and the $g$-th RIS element follows the Rician fading model as
\begin{equation}
{h}_{g,m} = \sqrt{\frac{\kappa }{\kappa +1}}{h}_{g,m}^{(LOS)} + \sqrt{\frac{1}{\kappa +1}}{h}_{g,m}^{(NLOS)},
\end{equation}
where $\kappa$ is the Rician factor, ${h}_{g,m}^{(LOS)}$ and ${h}_{g,m}^{(NLOS)}$ represent the line-of-sight (LOS) and non-line-of-sight (NLOS) parts of the fading channel, respectively. The LOS part can be specifically expressed as

\begin{equation}
{h}_{g,m}^{(LOS)}= \sqrt{{\beta}_{0}}{e}^{-j(g-1)\pi \text{sin}\theta}{d}_{g,m}^{-\frac{\alpha_1}{2}},
\end{equation}
where $\theta$ is the angle of arrival (AoA) of the signal at the RIS, ${\beta}_{0}$ represents the path loss factor at the reference distance of one meter, ${d}_{g,m}$ is the distance, and $\alpha_1$ is the path loss exponent for LOS. The NLOS part can be specifically expressed as
\begin{equation}
h_{g,m}^{(NLOS)} = \tilde{h}_{g,m}{d}_{g,m}^{-\frac{\alpha_2}{2}},
\end{equation}
where $\tilde{h}_{g,m}$ is the complex Gaussian small-scale fading with zero mean and unit variance, ${d}_{g,m}$ is the distance between $m$-th BS and the $g$-th element, and $\alpha_2$ is the path loss exponent for NLOS. Similarly, the channel coefficient between user ${V}_{n}$ and the $g$-th element of RIS is the same as above and can be expressed as ${h}_{n,g}$. Moreover, $\mathbf{\Theta}$ is the phase shift matrix of RIS, and its expression is
\begin{equation}
\mathbf{\Theta} = \text{diag}(a_1 e^{j \theta_1}, \ldots, a_g e^{j \theta_g}, \ldots, a_G e^{j \theta_G}),
\end{equation}
where ${a}_{g}$ and ${\theta}_{g}$ are the amplitude and phase shift of the $g$-th reflecting element of RIS, respectively. In this paper, we assume the RIS practical phase shift model with $R$-bit quantization and there are $2^R$ possible phase shifts \cite{9115725}.

The channel of this model conforms to the quasi-static Rayleigh fading channel and is modeled as
\begin{equation}
    h_{m,n} = \frac{\varepsilon }{\sqrt{d_{m,n}^{\alpha_3 } } } ,
\end{equation}
where $\varepsilon $ represents Gaussian distribution, ${d}_{m,n}$ is the distance between ${K}_{m}$ and ${V}_{n}$, and $\alpha_3$ is the path loss exponent in the Rayleigh fading model. The signal transmitted by ${K}_{m}$ is ${x}_{m,n}$, and the received signal at ${V}_{n}$ is
\begin{equation}
    {y}_{n}=\sqrt{P_m}(h_{m,n} + h^{H}_{n,g} \Theta h_{g,m}) x_{m,n} + \omega_0,
\end{equation}
where ${P}_{m}$ is the power at ${K}_{m}$,  $H$ is the conjugate transpose, and ${\omega}_{0}$ is the noise received at ${V}_{n}$. Then the SNR can be expressed as
\begin{equation}
    {SNR}_{m,n} = \frac{P_m |h^{H}_{n,g} \Theta h_{g,m}|^2}{N_0},
\end{equation}
where ${N}_{0}$ is the noise power at ${V}_{n}$. Assume that different users are on different subcarriers, so the interference between users is not considered. The transmission rate between ${K}_{m}$ and ${V}_{n}$ can be calculated as
\begin{equation}
    {R}_{m,n} = B_{m,n} \log_2 \left( 1 +  {SNR}_{m,n}\right),
\end{equation}
where ${B}_{m,n}$ is the bandwidth of the transmission link between ${K}_{m}$ and ${V}_{n}$.

The energy consumption of ${K}_{m}$ at a given time slot $t$ (with the length of each time slot represented by $\tau$) is determined by both the power used and the duration spent in that state, defined as
  \begin{equation}
\begin{aligned}
E_{\tau}(t) =  {\sum_{i=1}^{L}} P_{s_i} \tau_{s_i} + \sum_{j=1}^{L-1}P_{s_j} \tau_{s_j, s_{j+1}}+ \sum_{o=2}^{L}P_{s_o} \tau_{s_o, s_{o-1}},
\end{aligned}
\end{equation}
where $s_i$ is the SM state for $K_m$, and $s_{j+1}$, $s_{o-1}$ represent the SM states that follow $s_j$ and $s_o$, respectively. The transition from $s_j$ to $s_{j+1}$ indicates deactivation, while the transition from $s_o$ to $s_{o-1}$ indicates reactivation. $\tau_{s_i}$ represents the duration in SM state $s_i$, $\tau_{s_j, s_{j+1}}$ represents the deactivation time from $s_j$ to $s_{j+1}$, and $\tau_{s_o, s_{o-1}}$ represents the reactivation time from $s_o$ to $s_{o-1}$. The sum of $\tau_{s_i}$, $\tau_{s_j, s_{j+1}}$, and $\tau_{s_o, s_{o-1}}$ is equal to the length $\tau$ of a time slot $t$. The first term represents the power consumption in the SM state $s_i$ with the power $P_{s_i}$, the second term represents the power consumption during the transition from $s_j$ to $s_{j+1}$, where $P_{s_j}$ is the power in the starting SM state for the transition, and the third term represents the power consumption during the transition from $s_o$ to $s_{o-1}$. Then the energy consumption model of ${K}_{m}$ can be defined as
\begin{equation}
    {E}_{{K}_{m}}(t) = \sum_{t=1}^{T}E_{\tau}(t)
\end{equation}

The total energy consumption model of the system can be defined as
\begin{equation}
\begin{split}
    E_{\text{system}}(t) = \sum_{m=1}^{M} E_{{K_m}}(t),
\end{split}
\end{equation}
where $M$ is the number of BS in the system.

\subsection{Problem Formulation}
The optimization problem in this paper is to minimize the total energy consumption of the BSs cooperative through cell zooming under the delay constraint, which is a long-term optimization problem. Assuming that there are $T$ time slots in total, it is reasonable to minimize the average energy consumption. In this paper, we assume that the traffic load follows the characteristics of FTP mode 3, which is a specific traffic load model for generating traffic load, particularly suitable for cellular network environments \cite{3gpp_tr38913}. In this mode, the traffic load is generated with a fixed file size and the inter-arrival time follows a Poisson distribution, which can simulate burst and spike traffic patterns. In actual wireless communications, FTP Mode 3 traffic patterns account for a large proportion of certain application scenarios, such as online games, audio and video streaming. Therefore, we choose to use FTP Mode 3 to simulate the generation of traffic load to be closer to the real communication environment. The optimization problem formula is as follows

\begin{subequations}
\label{formulationProblem}
\begin{align}
\tag{12} \label{p1} &(P1):  \quad \min\limits_{{\mathcal{S}}_{M}(t),\mathcal{C}_M(t),\mathcal{Z}_{M,N}(t)}\sum_{t =1}^{T}{E}_{system}(t) \\
\label{p2}  &\quad\quad \text{s.t.} \quad  \mathcal{S}_M(t) = \{S_1, S_2, ..., S_M\}, \\
\label{p3}  & \quad \quad \quad \quad  \mathcal{C}_M(t) \in \{C_1, C_2, ..., C_M\},\\
\label{p5}  & \quad \quad \quad \quad \mathcal{Z}_{M,N}(t) \in \{Z_{1,1}, Z_{1,2}, ..., Z_{M,N}\}, \\
\label{p6}  & \quad \quad \quad \quad \sum_{m=1}^{M} Z_{m,n} \leq 1, \quad \forall n, \\
\label{p8}  & \quad \quad \quad \quad  \sum_{n=1}^{N} Z_{m,n} \leq N, \quad \forall m, \\
\label{p7}  & \quad \quad \quad \quad \mathcal{D}_{i}(t) <\mathcal{D}_{\max},
\end{align}
\end{subequations}
where the constraint in \eqref{p2} represents the SM state for each BS that $\mathcal{S}_{M}(t)$ is the set for all BSs; the constraint in \eqref{p3} represents the cell zooming level for each BS and $\mathcal{C}_{M}(t)$ is the set for all BSs; the constraint in \eqref{p5} represents the user association between ${K}_{m}$ and ${V}_{n}$, where 0 means that ${K}_{m}$ and ${V}_{n}$ are not connected, and 1 means ${K}_{m}$ and ${V}_{n}$ are connected with each other; \eqref{p6} means that a user can be served by at most one BS; \eqref{p8} means one BS can serve multiple users; \eqref{p7} represents the delay constraint; here, $\mathcal{D}_{j}(t)$ denotes the delay, which is the time between the arrival of traffic data $j$ at the BS and its complete transmission to the corresponding user. The delay must not exceed the maximum threshold, $\mathcal{D}_{\max}$.

The above problem is a representative example of BS cooperation EE. Due to the need for long-term optimization and the dynamic changes in channel states and traffic loads, traditional optimization methods such as convex optimization have high computational complexity, making them especially unsuitable for dealing with highly dynamic situations such as fading with small-scale changes. In contrast, deep learning is a pre-trained method that has lower prediction complexity and can adapt to dynamic environments through training, so we propose to use deep learning algorithms as the solution for the optimization problem in \eqref{p1}. In this paper, we first model the energy consumption optimization problem of the system as an MDP and use the proposed DRL approach to optimize the SM, cell zooming and user association. Second, we address the optimization of RIS reflection coefficients by employing a low-complexity optimization algorithm, DCCN, aiming to enhance the signal and maximize energy-savings.

\section{Energy-Efficient Algorithms Based on Deep Neural Networks: DRL and DCCN}
\label{sectionalgori}
In this paper, we first use DRL to optimize variables such as SM, cell zooming, and user association, and then use DCCN to optimize the phase of RIS. The PPO algorithm was introduced in \cite{schulman2017proximal} as a promising DRL algorithm. It is a policy-based method and uses the clip parameter to limit the range of policy updates, thereby bridging the gap between new and old policies, and offering better stability. In addition, multiple batches and episodes also contribute to the stability of training. The DCCN algorithm was introduced in \cite{huang2022machine}. It is a deep learning-based approach that trains directly using optimization objectives, thus sidestepping the dimensionality challenges typical of conventional reinforcement learning methods. In this section, we propose using an advanced PPO and DCCN algorithm to optimize the system energy consumption under the delay constraint. First, the system is modeled as an MDP, and then PPO is utilized to solve the MDP. Furthermore, we introduce a DCCN to further enhance the energy-saving performance by optimizing the RIS reflection coefficients to enhance signal and speed up the transfer rate.

\subsection{MDP Components of the Proposed System}
\begin{figure}[t!]
\centering
\includegraphics[width = 0.65\columnwidth]{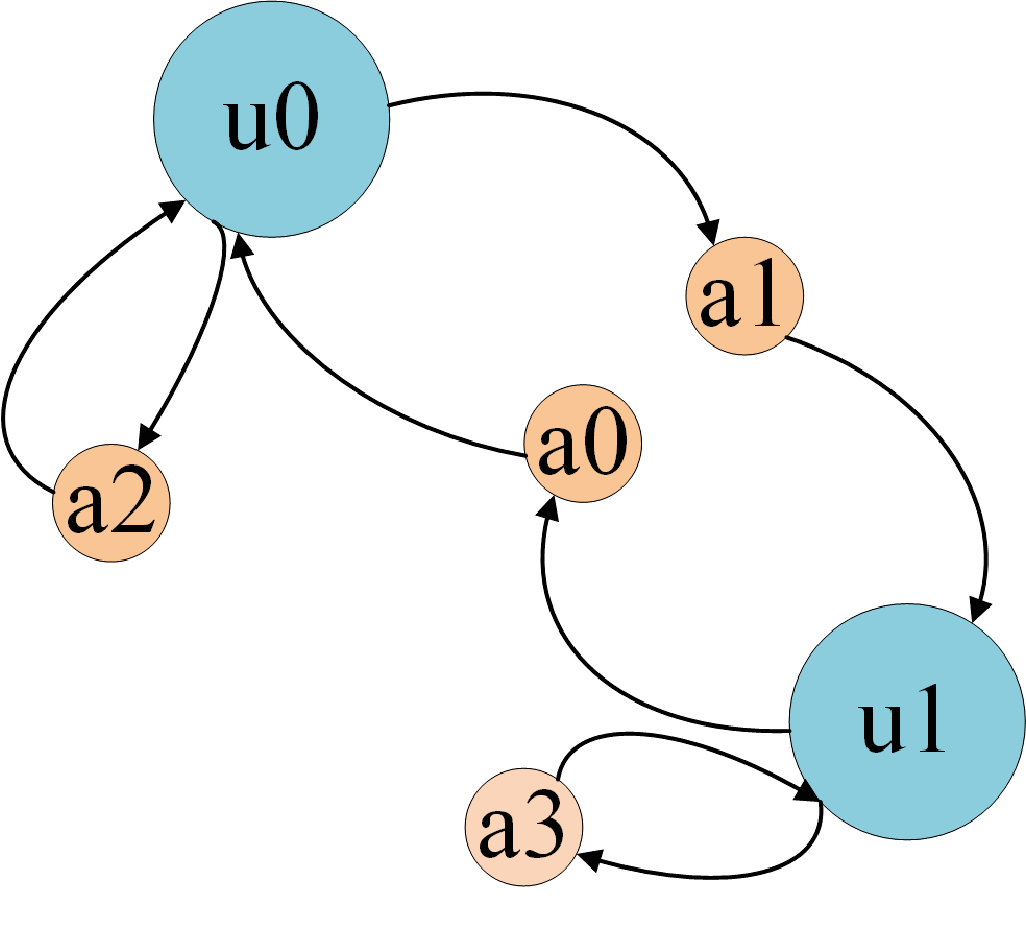}
\caption{MDP model.}
\label{MDP_change}
\end{figure}

The SM, cell zooming, and user association can be adjusted in each time slot, and the decision in the $t$-th time slot will be affected by the $(t-1)$-th time slot. Therefore, MDP can be represented as a tuple $\left (\mathcal{U}, \mathcal{A}, R\right )$, where $\mathcal{U}$ represents the possible system state and $\mathcal{A}$ represents the possible action set, as shown in Fig. \ref{MDP_change}.
$R$ is the reward function. The elements of MDP are introduced as follows:

1) State Set: The state set is denoted as $\mathcal{U}$ and the system state at time slot $t$ is $u(t)$, and it is composed of the pending traffic load, channel coefficients, and the SM states. The state at time slot $t$ conforms to the following expression
\begin{equation}
u(t) = \left (p(t), m(t),s(t), t\right ),
\end{equation}
where $p(t)$ represents the pending traffic load. $m(t)$ represents the channel coefficient matrix, $m(t) \in \mathbb{C} ^{M \times  N}$. $s(t)$ represents the SM state, $s(t) \in ^{M \times 1}$. $t$ is the current time slot.

2) Action Set: The action set is denoted as $\mathcal{A}$ and the action set at the time slot $t$ is $a(t)$ that is composed of the SM states, cell
 zooming levels and user association between ${K}_{m}$ and ${V}_{n}$. The action at time slot $t$ can be expressed as
\begin{equation}
a(t) = \left (s(t), c(t),z(t)\right ),
\end{equation}
where $s(t)$ represents the SM state in the action set, $s(t) \in \mathbb{C}^{M \times 1}$, $c(t)$ represents the cell zooming level, $c(t) \in \mathbb{C}^{M \times 1}$, $z(t)$ represents the association between ${K}_{m}$ and ${V}_{n}$, $z(t) \in \mathbb{C}^{M \times N}$. These variables are considered to be the optimization variables in \eqref{p1}.

3) Reward Function: We aim to find the action set that minimizes the total energy consumption for the proposed system. The total reward of the system at time slot $t$ consists of the energy consumption and delay of all BSs. It can be defined according to \eqref{p1}
\begin{equation}
\begin{aligned}
R_{u,a}(t) & = \mathcal{H} ( \sum_{t=1}^{T} \sum_{n=1}^{N} x(t))^{*} \left( L_{1} \mathcal{H}( P - P_{SM1} \right)^{*} \\
& \quad + L_{2} \mathcal{H}( P_{SM1} - P)^{*}) \\
& \quad + ( 1 - \mathcal{H} ( \sum_{t=1}^{T} \sum_{n=1}^{N} x(t))^{*}) \\
& \quad( L_{3} \mathcal{H} \left( \varrho (t) - \mathcal{D}_{\max} \right)^{*} + L_{4} \mathcal{H} \left(\mathcal{D}_{\max} - \varrho (t) \right)^{*}),
\end{aligned}
\end{equation}
where \(x(t)\) represents the pending traffic load in each time slot; \(P\) is the system energy consumption in the current time slot; \({P}_{\text{SM1}}\) is the energy consumption in the SM1 state;
\(\mathcal{D}_{\max}\) denotes the maximum delay constraint, and \(\varrho(t)\) is the delay at time slot \(t\).
\(\mathcal{H}[\cdot]^*\) is the Heaviside function, where \([\cdot]^* < 0\) results in \(\mathcal{H}[\cdot]^* = 0\), and \([\cdot]^* \geq 0\) results in \(\mathcal{H}[\cdot]^* = 1\).  \({L}_{1} \sim {L}_{4}\) represent the reward coefficients under different conditions. Specifically, \({L}_{1}\) and \({L}_{2}\) are energy-related coefficients, while \({L}_{3}\) and \({L}_{4}\) are delay-related coefficients. To prioritize delay constraints, \({L}_{3}\) and \({L}_{4}\) are given higher weights than \({L}_{1}\) and \({L}_{2}\). For example, when the delay constraint is violated, the penalty linked to the delay-related coefficients is set significantly higher than any potential energy-saving rewards. This ensures that the agent avoids actions that violate the delay constraint. Once the delay falls within a safe range, the agent then focuses on minimizing system energy consumption. Since PPO aims to maximize the cumulative reward, the above design for reward function allows energy saving while satisfying delay constraints, treating system optimization as a long-term objective. The first part of reward function $\mathcal{H} ( \sum_{t=1}^{T} \sum_{n=1}^{N} x(t) )^{*} \left( L_{1} \mathcal{H}( P - P_{SM1} \right)^{*} + L_{2} \mathcal{H}( P_{SM1} - P)^{*})$ is related to system energy consumption, which is used to minimize the system's energy consumption. This part works as follows: when the pending traffic load is zero, the value of the first Heaviside function is 1, and the system will reward actions that can keep the system energy consumption at a low level. The coefficients $L_1$ and $L_2$ are used to adjust the intensity of the reward or penalty according to the current energy state $P$ and the energy consumption $P_{SM1}$ in SM, thereby encouraging the system to choose actions that reduce energy consumption and give priority to low-power states when possible. The second part of reward function $( 1 - \mathcal{H} ( \sum_{t=1}^{T} \sum_{n=1}^{N} x(t)))^{*}( L_{3} \mathcal{H} \left( \varrho (t) - \mathcal{D}_{\max} \right)^{*} + L_{4} \mathcal{H} \left(\mathcal{D}_{\max} - \varrho (t) \right)^{*})$ is related to the delay constraint, which is used to ensure that the delay is always below the maximum allowed delay. It works like this: when the pending traffic load is non-zero, the delay constraint part is activated. If the delay exceeds the maximum delay constraint, a penalty is given; if it does not, a reward is given.

\subsection{DRL-Based Optimization Algorithm}

\begin{figure}[t!]
\centering
\includegraphics[width = \columnwidth]{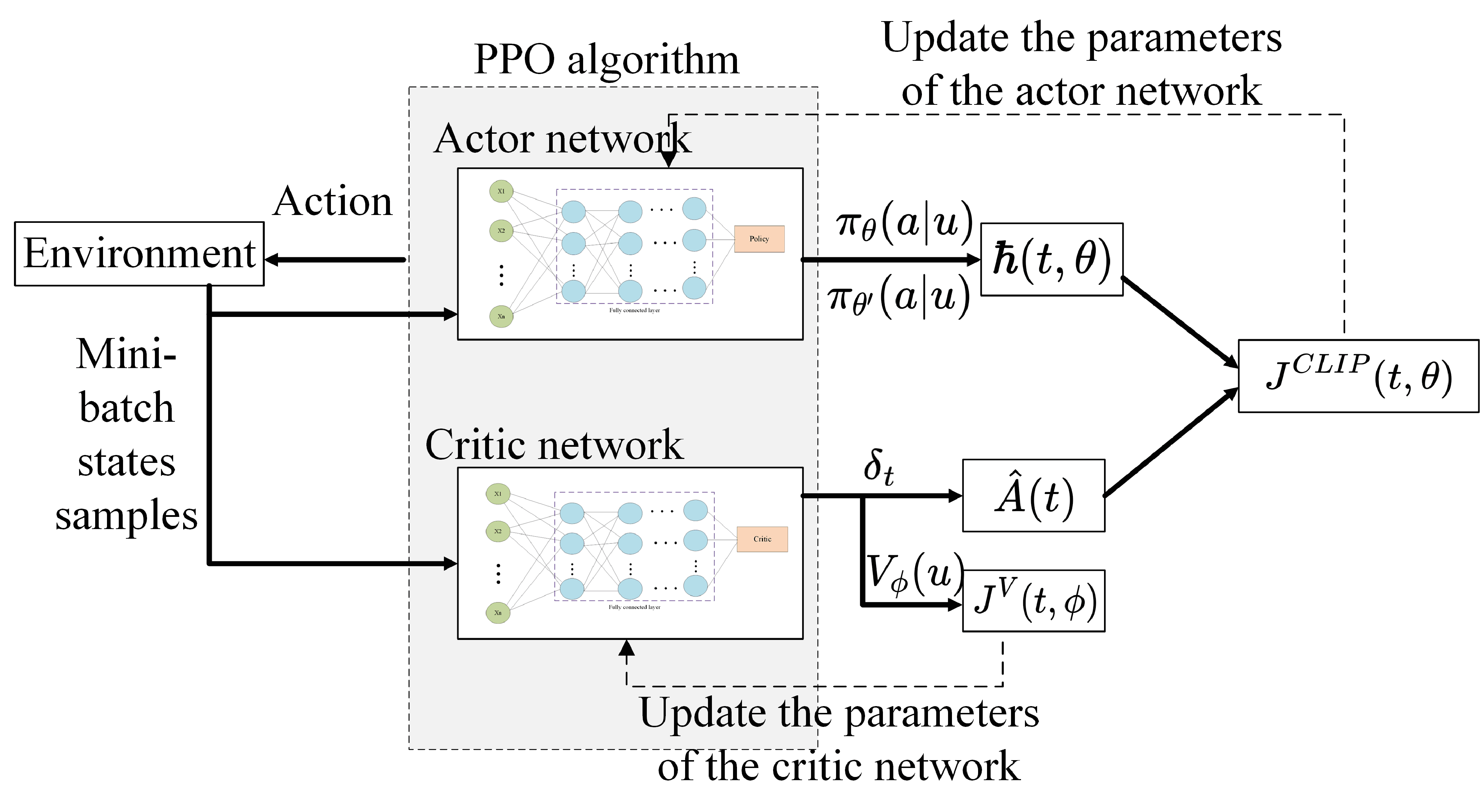}
\caption{PPO algorithm structure.}
\label{PPO_struc}
\end{figure}

To optimize the variables for \eqref{p1} using the MDP framework, we employ PPO algorithm. This actor-critic algorithm consists of two neural networks: the policy (actor) network and the value (critic) network as shown in Fig. \ref{PPO_struc}, the output of the policy network is $\pi_{\theta}(a|u)$, which is a probability distribution of actions $a$ under a given state $u$, with $\theta$ as its network parameters. And the $\pi_{\theta'}(a|u)$ with the $\theta'$ means the old network parameters before update.  The output of the value network is $V_{\phi}(u)$ which estimates the expected reward for a state $u$ under policy $\pi$, with $\phi$ as its network parameters. The $\hat{A}(t)$ in Fig. \ref{PPO_struc} is the generalized advantage estimator (GAE).
Both networks use fully connected layers with the ReLU activation function. The output of the policy network is converted to a probability distribution using the softmax function, while the value network outputs a scalar state value estimate. The value network is updated by minimizing the mean squared error (MSE) between the temporal difference target (TD) and its output \cite{9501950}, which can be expressed as
\begin{equation}
\begin{split}
     J^{V}(t,\phi) &= \hat{\mathbb{E}}[(V_{\phi}(u(t)) - \mathcal{T}(t))^2],
\end{split}
\label{citirc}
\end{equation}
where  \( \hat{\mathbb{E}}[\cdot] \) denotes the expectation, $u(t)$ is the state at time slot $t$, ${V}_{\phi }(u(t))$ is the state value function, ${T}(t)$ is the TD target and the calculation for each time slot is as follows
\begin{equation}
\mathcal{T}(t)=r(t+1)+\gamma {V}_{\phi }(u(t+1))(1-d(t)),
\end{equation}
where $r(t+1)$ is the immediate reward received after taking the action at time slot $t$, $u(t+1)$ is the state as time slot $t+1$, $\gamma$ is the discount factor that affects the current value of future rewards, ${V}_{\phi }(u(t+1))$ is the reward prediction for the next state $u(t+1)$, and $d(t)$ indicates whether it is an end state. Building upon \eqref{citirc}, the evaluation of the current prediction performance of the value function and the guidance for its update are encapsulated by the TD residual, expressed as follows
\begin{equation}
\delta(t) = r(t) + \gamma V(u(t+1)) - V(u(t)),
\end{equation}
where $V(u(t))$ is the value estimate of the current state $u(t)$, and $V(u(t+1))$ is the state value function for $u(t+1)$. The policy network is updated by optimizing a clipped surrogate objective function. This involves minimizing the negative expected value of the minimum between two terms: the product of the probability ratio and the GAE \cite{schulman2017proximal}, and the product of the clipped probability ratio and the GAE. This approach encourages policy improvement within a defined range, ensuring stability and avoiding excessively large updates. The policy optimization function can be expressed at
\begin{equation}
\begin{split}
     J^{CLIP}(t,\theta) &= \hat{\mathbb{E}}[ \min(\hbar(t,\theta) \hat{A}(t), \\ & \text{clip}(\hbar(t,\theta), 1-\epsilon, 1+\epsilon) \hat{A}(t)) ],
\label{jclip}
\end{split}
\end{equation}
where \( \hat{\mathbb{E}}[\cdot] \) denotes the expectation,
${\hbar}(t,\theta)$ is the probability ratio between the old policy and the new policy, $\theta$ is the parameter of the network, and $\hat{A}(t)$ is the GAE, $\epsilon$ is the hyperparameter controlling the clip range, $\epsilon \in [0,1]$. To avoid instability in learning caused by large policy updates from $\theta$ to $\theta^{'}$, a constraint ${\hbar}(t,\theta)$ can be defined within the interval [1-$\epsilon$, 1+$\epsilon$]
\begin{equation}
\text{clip}(\hbar(t,\theta), 1-\epsilon, 1+\epsilon)=\left\{\begin{matrix}1-\epsilon,
 & \text{if}\quad{\hbar}(t,\epsilon )\leqslant 1-\epsilon,\\ 1+\epsilon,
 & \text{if}\quad{\hbar}(t,\epsilon )\geqslant  1+\epsilon,\\ {\hbar}(t,\epsilon ),
 & \textnormal{otherwise,}
\label{clip}
\end{matrix}\right.
\end{equation}
The expression of ${\hbar}(t)$ for \eqref{jclip} is as follows
\begin{equation}
{\hbar}(t,\theta)=\frac{{\pi }_{{\theta}^{'}}(a|u)}{{\pi }_{\theta }(a|u)},
\label{action ratio}
\end{equation}
where ${\pi}_{{\theta}^{'}}(a|u)$ is the probability of taking action $a$ in state $u$ under the new policy, ${\theta }^{'}$ is the parameter of the new policy network, ${\pi}_{\theta}(a|u)$ is the probability of taking action $a$ in state $u$ under the old policy, ${\theta }$ is the parameter of the old policy network. To improve sample efficiency and increase the stability of learning, we consider a clipped version of the GAE for \eqref{jclip}, which can be expressed as follows
\begin{algorithm}[t!]
\caption{Proximal Policy Optimization (PPO) Training Process}\label{Algorithm 1}
\begin{algorithmic}[1]
\State \textbf{Initialization:} Initialize policy network $\pi_{\theta}$ with parameters $\theta$ and value network $V_{\phi}$ with parameters $\phi$
\State \textbf{Set:} Set hyperparameters: number of iterations $I$, steps per iteration $\mathcal{L}$, minibatch size $\mathcal{M}$, epochs $E$, discount factor $\gamma$, GAE parameter $\lambda$, PPO clipping parameter $\epsilon$
\For{$iteration = 1, 2, \ldots, I$}
    \State Initialize an empty trajectory buffer $\mathcal{B}$
    \For{$step = 1, 2, \ldots, \mathcal{L}$}
        \State Observe state $u(t)$ and select action $a \sim \pi_{\theta}(a|u)$
        \State Execute action $a(t)$ in the environment
        \State Observe next state $u(t)$, reward $r(t)$, and whether \Statex\quad\quad\quad done $d(t)$
        \State Store transition $(u(t), a(t), r(t), u(t+1), d(t))$ in $\mathcal{B}$
        \If{$d(t)$ is True}
            \State Reset the environment
        \EndIf
    \EndFor
    \State Compute advantage estimates $\hat{A}$ using $\mathcal{B}$ with GAE
    \State Compute discounted returns $R_{u,a}$ for each step in $\mathcal{B}$
    \For{$epoch = 1, 2, \ldots, E$}
        \For{each minibatch $mb$ of size $\mathcal{M}$ sampled from \Statex\quad\quad\quad$\mathcal{B}$}
            \State Compute policy ratio according to \eqref{action ratio}.
            \State Calculate clipped objective according to \eqref{clip}.
            \State Compute clipped surrogate objective $J^{\text{CLIP}}(t,\theta)$ \Statex \quad\quad\quad\quad\quad according to \eqref{jclip}.
            \State Compute value function loss $J^{\text{V}}(t,\phi)$ according \Statex \quad\quad\quad\quad\quad to \eqref{citirc}.
            \State Update policy by maximizing $J^{\text{CLIP}}(t,\theta)$ w.r.t. \Statex \quad\quad\quad\quad\quad $\theta$
            \State Update value network by minimizing $J^{\text{V}}(t,\phi)$\Statex \quad\quad\quad\quad\quad w.r.t. $\phi$
        \EndFor
    \EndFor
    \State $\theta_{\text{old}} \gets \theta$
\EndFor
\State \textbf{return} optimized policy $\pi_{\theta}$
\end{algorithmic}
\end{algorithm}
\begin{equation}
\hat{A}^{GAE(\gamma, \lambda)}(t) = \sum_{k=0}^{\infty} (\gamma \lambda)^k \delta(t+k),
\end{equation}
where $\gamma$ is the discount factor, $\gamma \in [0,1]$, $\lambda$ is a parameter used to trade off bias and variance that can effectively control the balance between exploration and utilisation in the PPO learning process, $\delta(t+k)$ is the TD residual of time step $t+k$. The overall maximisation goal of PPO is as follows according to \eqref{citirc} and \eqref{jclip}
\begin{equation}
{J}^{PPO}(\theta ,\phi )= \hat{\mathbb{E}}[ {J}^{CLIP}(t,\theta)-q{J}^{V}(t,\phi )],
\end{equation}
where $q$ is the loss coefficient. The pseudo-code of the PPO-based optimization algorithm is summarized in Algorithm \ref{Algorithm 1}. First, the policy (actor) network and the value (critic) network are initialized, and the relevant hyperparameters are set. In each iteration, the agent interacts with the environment, generating and storing trajectory data that includes the state, action, reward, and next state. Then the advantage function is then computed using GAE function, and the policy and value functions are optimized through the clipping mechanism of PPO. Finally, the parameters of both the policy and value networks are updated, allowing the model to be gradually refined.

\subsection{DCCN-Based Energy-Efficient Algorithm}

DRL is a widely used algorithm for solving optimization problems through the interaction between the agent and the environment. However, the optimization of RIS reflection coefficients presents a challenge for DRL due to the discrete phase shift, which results in high dimensionality. Additionally, the working mechanism of DRL relies on observing cumulative rewards during each training process, making it particularly suitable for long-term optimization goals.
\begin{figure}[t!]
\centering
\includegraphics[width = \columnwidth]{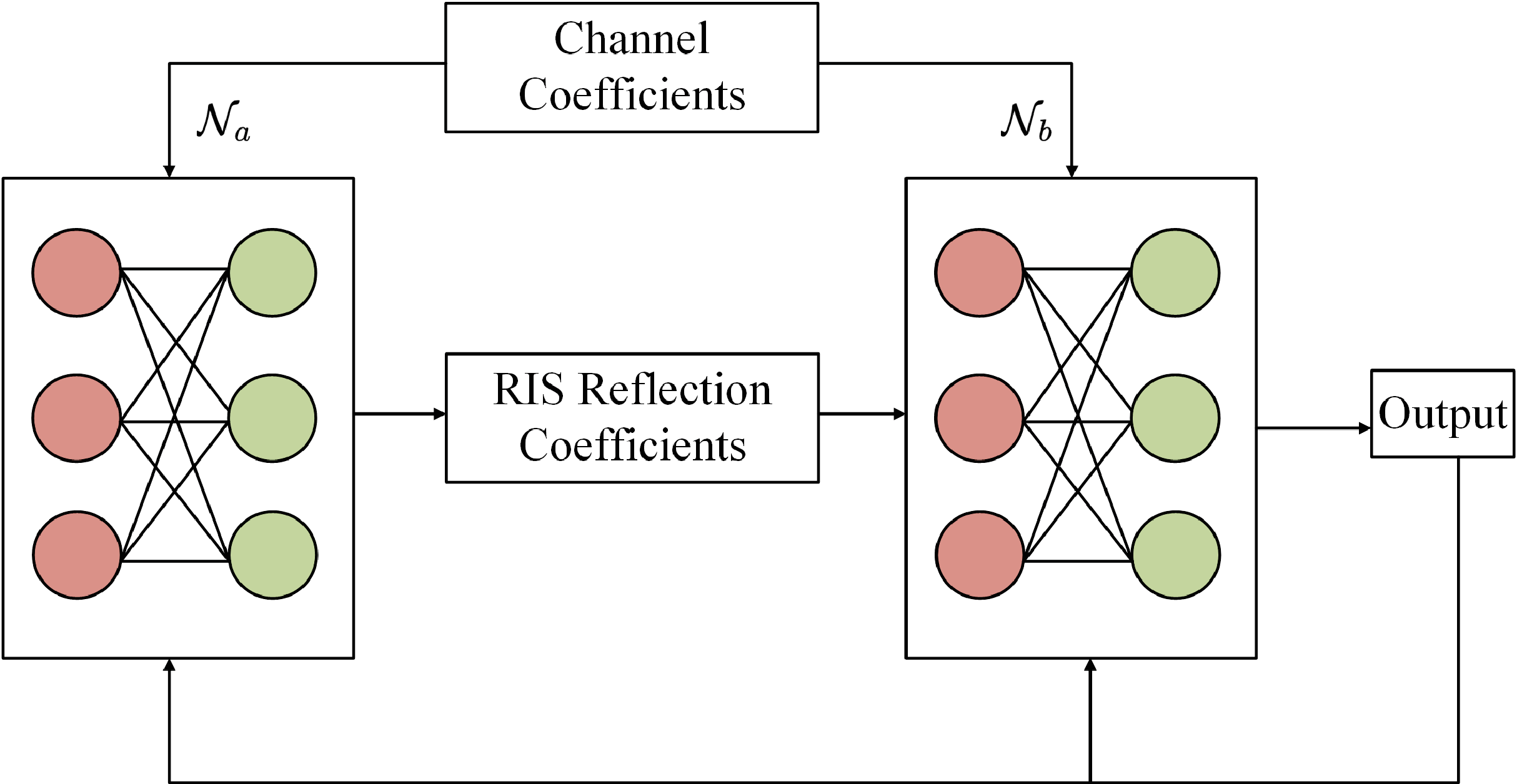}
\caption{DCCN algorithm structure.}
\label{DCCN_struc}
\end{figure}
However, during transmission, the channel coefficients are independent in each time slot. Hence, the optimization of RIS reflection coefficients needs to find the optimal solution in each time slot rather than relying on long-term accumulated rewards. To solve this problem, this paper proposes using a deep neural network ${\mathcal{N}}_{a}$ to optimize the RIS reflection coefficients. Since the RIS reflection coefficients are related to the channel state, we use the current channel state as the input of the neural network and the RIS reflection coefficients as the output, thereby directly obtaining the mapping relationship between them. The mapping relationship between channel matrices and RIS reflection coefficients can be expressed as
\begin{equation}
\mathbf{\Theta}= {\mathcal{N}}_{a}(\mathcal{H};W_a,b_a),
\end{equation}
where $\mathcal{H}$ is related channel matrices, $\mathbf{\Theta}$ is RIS reflection coefficients, $W_a$, and $b_a$ are the parameters of network ${\mathcal{N}}_{a}$ that $W_a$ is the weight and $b_a$ is bias.

However, during the training process of this neural network, some pre-labelled data is required to update the network parameters ($W_a$ and $b_a$) to obtain a more accurate mapping between channel matrices and RIS reflection coefficients. Here, the labels refer to the optimal RIS reflection coefficients corresponding to each specific channel state. Since the channel state of each link varies randomly and independently, collecting sufficient labelled data to cover all possible channel states is difficult, which makes it challenging to directly train the neural network to predict the optimal RIS reflection coefficients. Moreover, the optimal RIS reflection coefficients do not directly correlate with maximizing the transmission rate, which is crucial for transitioning the BS into sleep states. To address this problem, this paper proposes using DCCN \cite{huang2022machine}, the structure is shown in Fig. \ref{DCCN_struc}. DCCN consists two networks that are ${\mathcal{N}}_{a}$ network and ${\mathcal{N}}_{b}$ network. ${\mathcal{N}}_{a}$ is the network mentioned above, which takes the channel state as input and outputs the optimal RIS reflection coefficients. Since the overall goal is to maximize system energy efficiency by allowing the BS to enter sleep states as much as possible, we use the maximization of the transmission rate as the optimization objective of DCCN. The input of ${\mathcal{N}}_{b}$ is related channel coefficients, the user association matrix and the RIS reflection coefficients, the output of ${\mathcal{N}}_{a}$ (RIS reflection coefficients), the output of ${\mathcal{N}}_{b}$ is the channel capacity. The mapping relationship between channel matrices, RIS reflection coefficients and channel capacity can be expressed as
\begin{equation}
\mathcal C= {\mathcal{N}}_{b}(\mathbf{\Theta},\mathcal{H};W_b,b_b),
\end{equation}
where $W_b$ and $b_b$ are the parameters of network ${\mathcal{N}}_{b}$.

Both ${\mathcal{N}}_{a}$ and ${\mathcal{N}}_{b}$ update their network parameters through backpropagation during the training process. The parameters of ${\mathcal{N}}_{a}$ can be updated according to the gradient descent which can be expressed as
\begin{equation}
\mathbf{W}_{a,new} \leftarrow \mathbf{W}_a - \eta \frac{\partial L_a}{\partial \mathbf{W}_a},
\end{equation}
where $\eta$ is the learning rate, $\frac{\partial L_a}{\partial \mathbf{W}_a}$ is the gradient of ${\mathcal{N}}_{a}$, $\partial$ represents the partial derivative, $\mathbf{W}_{a,new}$ is the new network parameter after the update, and $\mathbf{W}_{a}$ is the old network parameter. The update of the bias is as follows:
\begin{equation}
\mathbf{b}_{a,new} \leftarrow \mathbf{b}_a - \eta \frac{\partial L_a}{\partial \mathbf{b}_a},
\end{equation}
where $\frac{\partial L_a}{\partial \mathbf{b}_a}$ is the gradient of ${\mathcal{N}}_{a}$ with respect to the bias $\mathbf{b}_a$, $\mathbf{b}_{a,new}$ is the new network bias after the update, and $\mathbf{b}_a$ is the old network bias. Since the goal of ${\mathcal{N}}_{b}$ is to maximize the channel capacity, we can use the channel capacity $\mathbf{C}_{m,n}$ between BS ${K}_{m}$ and user ${V}_{n}$ as the loss function. However, to optimize using gradient descent, we usually convert a maximization objective into a minimization objective. Thus, the loss function can be defined as the negative of the channel capacity
\begin{equation}
    L_b = -\mathbf{C}_{m,n}
\end{equation}

First, ${\mathcal{N}}_{b}$ is trained based on a given set of channel matrices and RIS reflection coefficients. The goal is to minimize the loss function $L_b$, where $L_b$ is defined as the negative of the channel capacity. Through backpropagation and gradient descent, the parameters ($W_b$ and $b_b$) of ${\mathcal{N}}_{b}$ are updated iteratively to minimize $L_b$. In this way, ${\mathcal{N}}_{b}$ learns to more accurately map the relationship between channel matrices, RIS reflection coefficients, and channel capacity. After ${\mathcal{N}}_{b}$ is trained, its network parameters are fixed, and the DCCN network training begins. During DCCN training, the network parameters ($W_a$ and $b_a$) of ${\mathcal{N}}_{a}$ are updated using backpropagation and gradient descent. At this stage, the RIS reflection coefficients in the DCCN network serve merely as intermediate variables. Once the DCCN network training is completed, ${\mathcal{N}}_{a}$ will be able to determine the optimal RIS reflection coefficients to accelerate data transmission rates, allowing the BS to enter sleep mode as quickly as possible to achieve energy-savings.

\section{Simulation And Results}
\label{sectionresult}
\label{section_simulation}

In this section, we analyze the performance of the proposed scheme through simulations. For comparison, we use a variety of experimental schemes as benchmarks. Unless otherwise stated, the parameters of the proposed communication system model are shown in Table \ref{T4.1}. In addition, the locations of the BSs and users are randomly generated. The locations of the three BSs are (148.24 m, 201.12 m), (107.99 m, 112.61 m) and (204.57 m, 124.73 m), and the locations of the five users are (147.03 m, 110.94 m), (140.98 m, 161.71 m), (188.24 m, 165.65 m), (199.17 m, 89.26 m), (149.17 m, 141.26 m), respectively. Table \ref{T4.2} presents the power consumption of BSs under different SM levels and cell zooming levels, the transition times between SMs, the hold times for each SM state, the coverage radius ranges under cell zooming are 40 m for zooming in, 50 m for no zooming, and 60 m for zooming out, the number of elements of RIS, and the power for each element of RIS. The duration of one time slot is 1 ms. We assume that the traffic load conforms to FTP mode 3. For the DRL, the proposed scheme utilizes an actor network and a critic network, each with hidden layer neurons set to 128. The learning rate of PPO and DQN is the same. For the DCCN, both network ${\mathcal{N}}_{a}$ and network ${\mathcal{N}}_{b}$ are fully connected networks, and $M$ is the number of neurons for the hidden layer. The remaining parameters are shown in Table \ref{T4.1}. In the simulations, we use PPO-optimized SM (PS), PPO-optimized cell zooming (PZ), PPO-optimized SM assisted by cell zooming (PSZ), a scheme without any energy-saving techniques and optimization algorithms (AA), and a DQN-optimized SM assisted by cell zooming and RIS (DSZR) as benchmarks.

\begin{figure}[t!]
\includegraphics[width=\columnwidth]{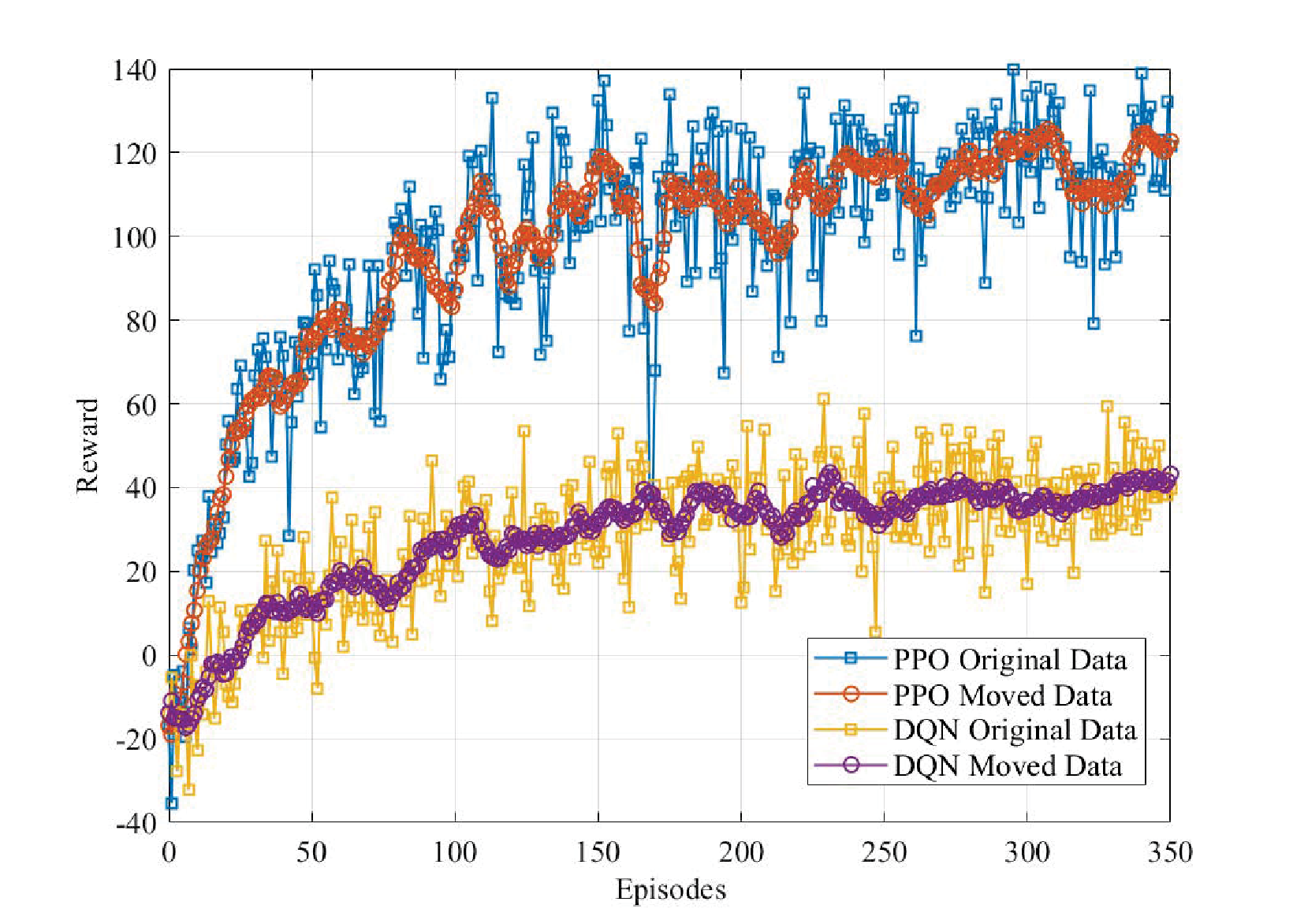}
\caption{Reward convergence of the proposed PPO-based energy-saving algorithm and the benchmark DQN-based
algorithm.}
\label{F4.10}
\end{figure}

To demonstrate the performance of the proposed PPO algorithm, DQN is used as a benchmark algorithm, which is a DRL algorithm based on Q-learning and utilizes deep neural networks (DNN) to approximate the Q-value function. It is one of the most classic and widely recognized methods in reinforcement learning. Fig. \ref{F4.10} illustrates the reward convergence for optimizations based on PPO and DQN. In these figures, the data have been smoothed using a moving average technique to reduce noise and better highlight the overall trend. It can be clearly seen from the results that the reward under the delay constraint increases with the number of iterations. The reward of the proposed PPO algorithm reaches about 117 after 150 iterations, while the reward of the DQN algorithm after 150 iterations is only about 30. This indicates that PPO can obtain better cumulative reward values than the DQN algorithm in the same task, demonstrating the superiority of PPO in terms of energy-savings for the system in different states under the same delay constraint. On the other hand, DQN shows greater volatility in reward values during training compared to PPO, indicating less stability. DQN optimizes the Q-value function, whereas PPO directly optimizes the policy, thereby avoiding the instability caused by approximating the Q-value function. Therefore, the simulations indicate that the PPO-based solution outperforms the DQN-based approach.
\begin{table}[!t]
\footnotesize
\centering
\caption{Parameters in Simulations.}
\label{T4.1}
\begin{tabular}{|l|l|}
\hline
\textbf{Parameter} & \textbf{Value} \\ \hline
Numbers of BSs: $M$ & 3 \\ \hline
Number of users: $N$ & 5 \\ \hline
Noise power: $N_0$ & -80 dBm \cite{huang2022machine} \\ \hline
Path loss exponent for LOS: $\alpha_1$ & 2 \\ \hline
Path loss exponent for NLOS: $\alpha_2$ & 3.5 \\ \hline
Path loss exponent for Rayleigh fading: $\alpha_3$ & 3.5 \\ \hline
$\beta_{0}$ & -10 dB \\ \hline
Rician factor: $\kappa$ & 10 dB \\ \hline
Bandwidth: ${B}_{m,n}$ & 20 MHz \\ \hline
Learning rate in PPO-DRL and DQN-DRL & 0.003 \\ \hline
Discount factor for PPO-DRL: $\gamma$ & 0.98 \\ \hline
Generalized advantage parameter for PPO-DRL: $\lambda$ & 0.95 \\ \hline
Clipping parameter for PPO-DRL: $\epsilon$ & 0.2 \\ \hline
Number of training iterations for PPO & 300 \\ \hline
Number of layers & 3 \\ \hline
Neurons for hidden layer & 128 \\ \hline
Learning rate & 0.001 \\ \hline
Coverage radius for zooming in & 40 m \\ \hline
Coverage radius for not zooming & 50 m \\ \hline
Coverage radius for zooming out & 60 m \\ \hline
Number of RIS elements: $G$ & 128 \\ \hline
Power for each RIS element: $P_g$ & 1.5 mW \cite{huang2019reconfigurable}  \\ \hline
Quantization level of RIS & 3 bit \\ \hline

\end{tabular}
\end{table}

\begin{figure}[t!]
\includegraphics[width=\columnwidth]{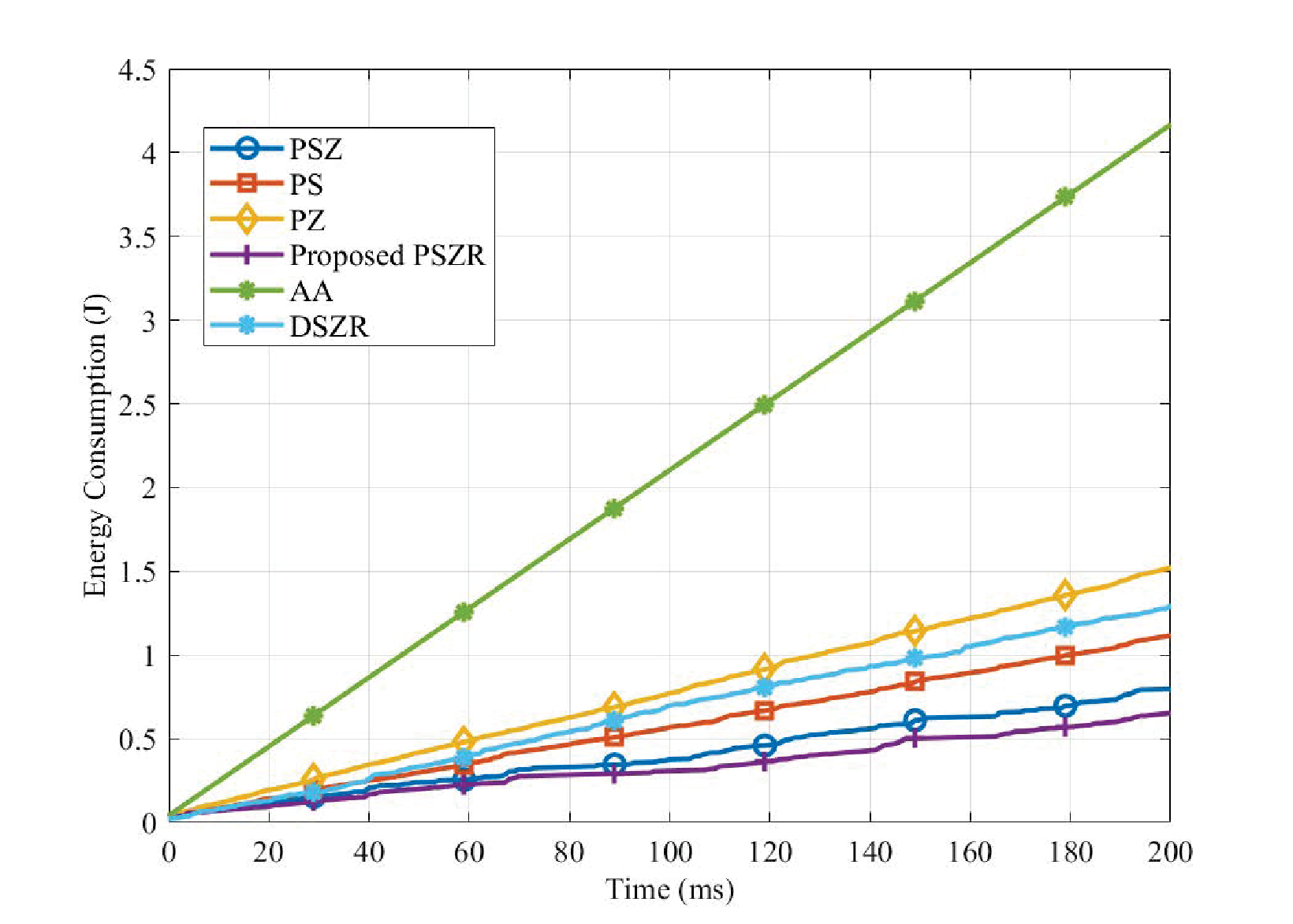}
\caption{The system energy consumption over time.}
\label{F4.4}
\end{figure}

Fig. \ref{F4.4} compares the performance of different energy-saving technologies. The average inter-arrival time of all traffic loads is set to 20 ms, and the size of the traffic load is 0.05 MB. Scheme PSZR is the proposed approach in which BSs use SM, cell zooming, and RIS. As can be seen from Fig. \ref{F4.4}, the energy consumption of the seven schemes from high to low is AA, PZ, DSPR, PS, PSZ, and PSZR which reflects the energy-saving capability of the proposed scheme PSZR. First, from the algorithm perspective, since PPO is a DRL algorithm based on policy gradient and incorporates the clip strategy to limit the updating of policy amplitude, it performs more stably in complex environments. In contrast, DQN demonstrates lower stability and reward convergence than PPO, resulting in generally higher energy consumption for the DQN-optimized DSZR scheme compared to the PPO-optimized PSZR scheme. Moreover, in terms of energy-saving techniques, the reason why the PSZR saves more energy than PSZ is that under the same circumstances, RIS can reconstruct the propagation environment of wireless signals through low-energy passive elements, enhancing the signal under the same transmit power, and thereby speeding up the data transmission rate. When transmitting the same amount of traffic load, PSZR enables the BS to complete the transmission task more quickly, allowing it to enter sleep states sooner and achieve greater energy-savings. Furthermore, the PSZ scheme has lower energy consumption compared to the PS scheme because, when the BS is in active mode, cell zooming provides more possibilities for the BS to enter the sleep state. When a certain BS is at a low traffic load level and its adjacent BSs are at a high level, the adjacent BSs can zoom out to take over the load of the low-load BS, thereby allowing the BS to enter a sleep state. When a BS is at a high load level and its adjacent BSs are at a low load level, the BS zooms in to balance the load with the adjacent BSs to complete the transmission task as quickly as possible and enter a sleep state. In addition, the PS scheme has lower energy consumption compared with PZ because the SM allows the BS to enter a low-power sleep state, thereby significantly reducing energy consumption during low traffic loads, while the BS of the PZ scheme is always in active mode, therefore the energy consumption is relatively high. Furthermore, the energy consumption increase rates of the four solutions PZ, PS, PSZ and PSZR decrease from largest to smallest. As time goes by, the energy consumption gap between the four solutions will become larger, which also reflects the effect of long-term optimization.

\begin{table*}[!t]\footnotesize
\centering
\caption{The SM and cell zooming parameters of BS \cite{7145603}}
\label{T4.2}
\begin{tabular}{|l|cccc|c|c|c|}
\hline
\multicolumn{1}{|c|}{\multirow{2}{*}{Mode}} & \multicolumn{4}{c|}{Activate} & \multirow{2}{*}{SM1} & \multirow{2}{*}{SM2} & \multirow{2}{*}{SM3} \\ \cline{2-5}
\multicolumn{1}{|c|}{} & \multicolumn{1}{c|}{Zooming out} & \multicolumn{1}{c|}{Not zooming} & \multicolumn{1}{c|}{Zooming in} & Idle & & & \\ \hline
Power (W) & \multicolumn{1}{c|}{7.3} & \multicolumn{1}{c|}{6.9} & \multicolumn{1}{c|}{6.6} & 2.3 & 1.5 & 0.4 & 0.3 \\ \hline
Transition times (ms) & \multicolumn{3}{c}{Fast transition} & & 0.071 & 1 & 10 \\ \hline
Hold time (ms) & \multicolumn{3}{c}{No hold time} & & 0.07 & 1 & 0 \\ \hline
\end{tabular}
\end{table*}

\begin{figure}[t!]
\includegraphics[width=\columnwidth]{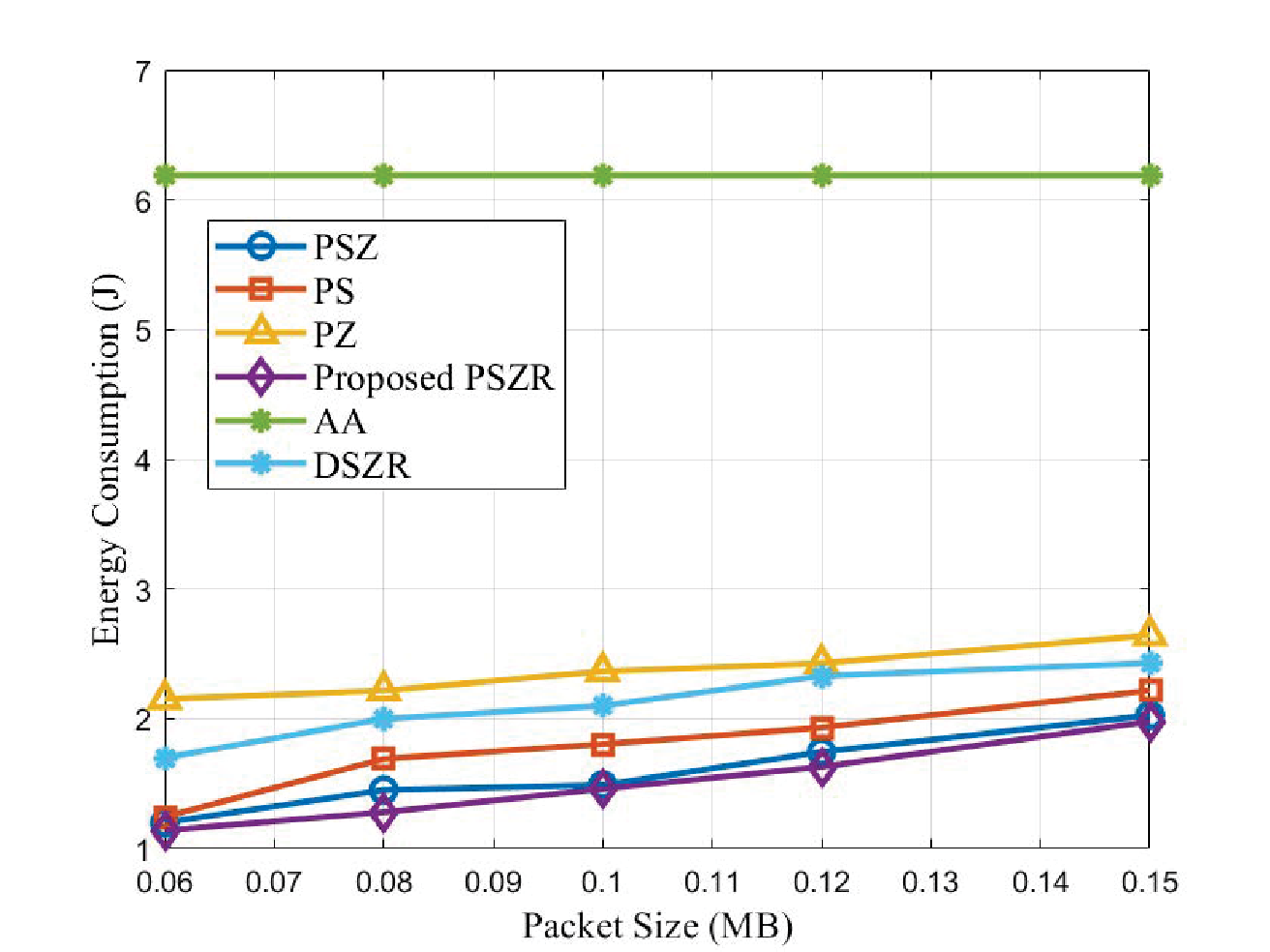}
\caption{The system energy consumption at different packet sizes.}
\label{F4.5}
\end{figure}

The effect of traffic load on the energy consumption of a system with the delay constraint is illustrated in Fig. \ref{F4.5}. Here, the average inter-arrival time of the traffic load is set to 30 ms. The figure shows traffic loads from 0.06 MB to 0.15 MB, respectively. This figure shows that for all schemes, when the packet size increases, the cumulative energy consumption of the system also increases at an accelerated rate. This is because the BS needs a longer time to complete the transmission task, which leads to a reduction in the time to enter a sleep state, which results in increased energy consumption. Furthermore, compared with the baseline AA, the proposed PSZR scheme significantly reduces system energy consumption under different data packet sizes.

\begin{figure}[t!]
\includegraphics[width=\columnwidth]{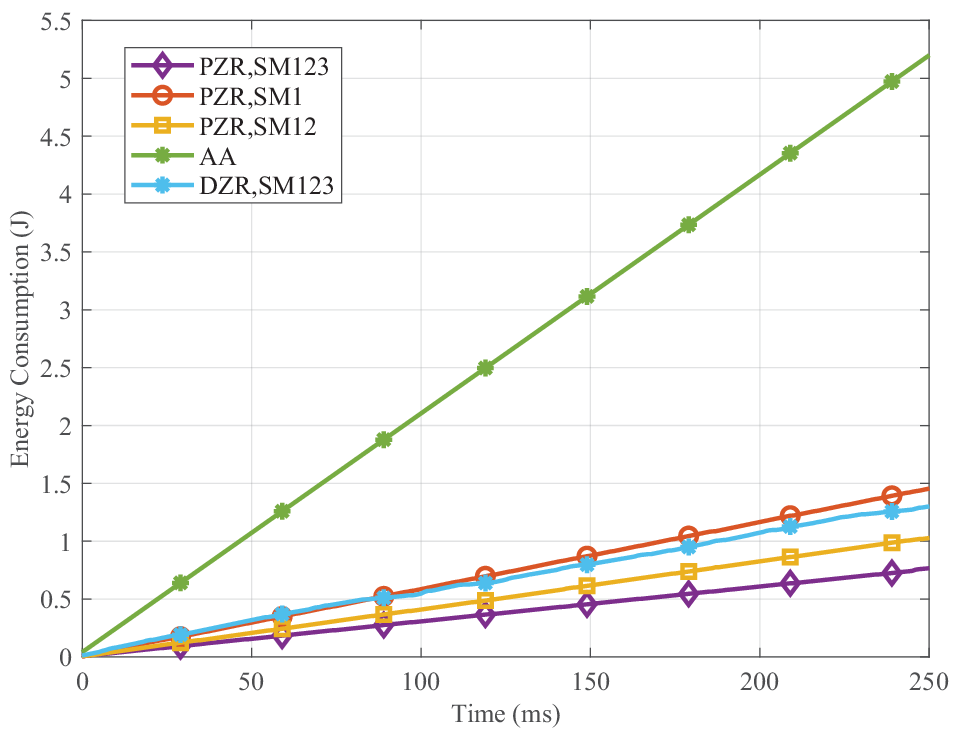}
\caption{The system energy consumption under different sleep levels.}
\label{F4.6}
\end{figure}

\begin{figure}[t!]
\includegraphics[width=\columnwidth]{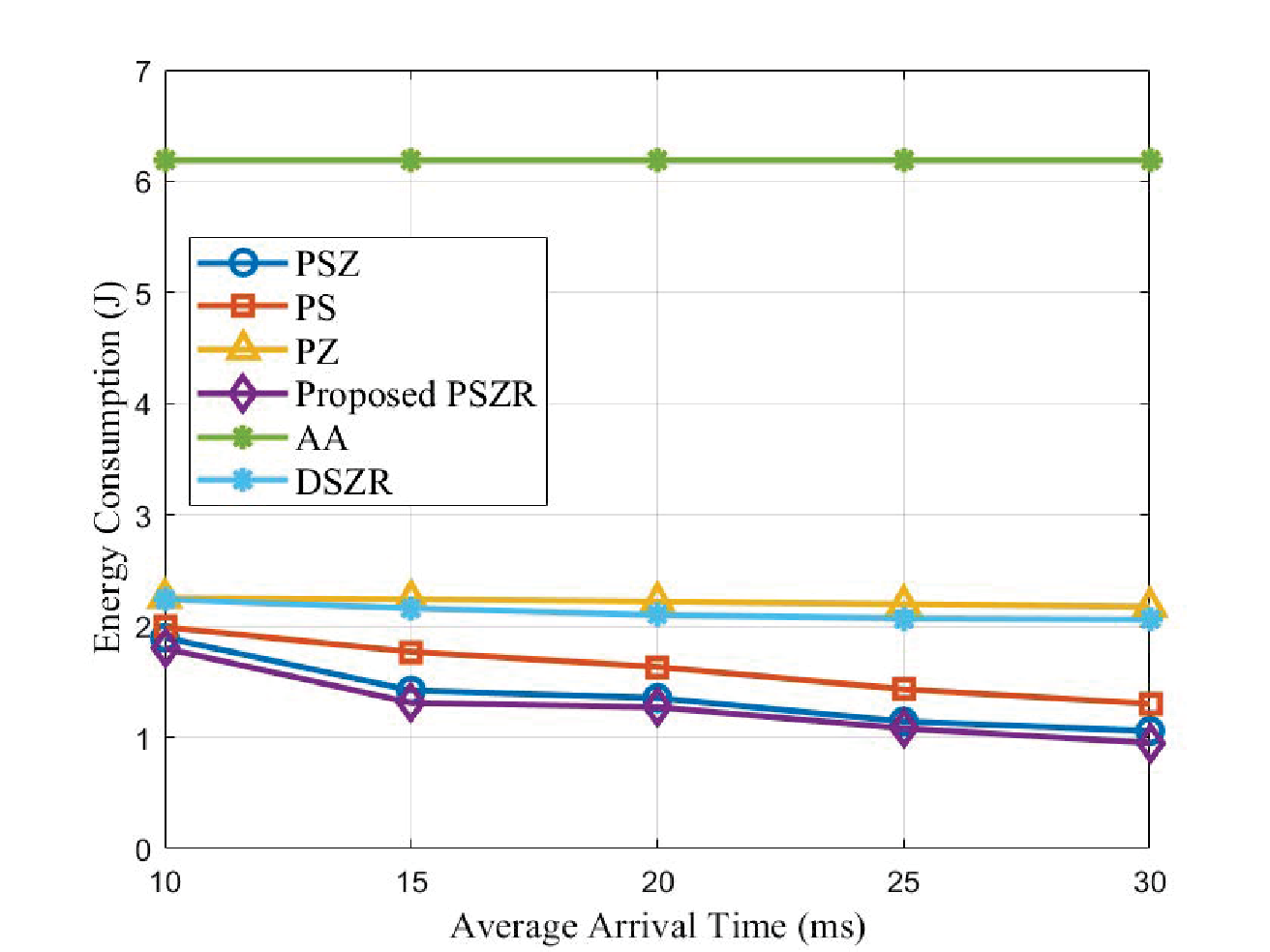}
\caption{The system energy consumption at different average inter-arrival times.}
\label{F4.7}
\end{figure}

\begin{figure}[t!]
\includegraphics[width=\columnwidth]{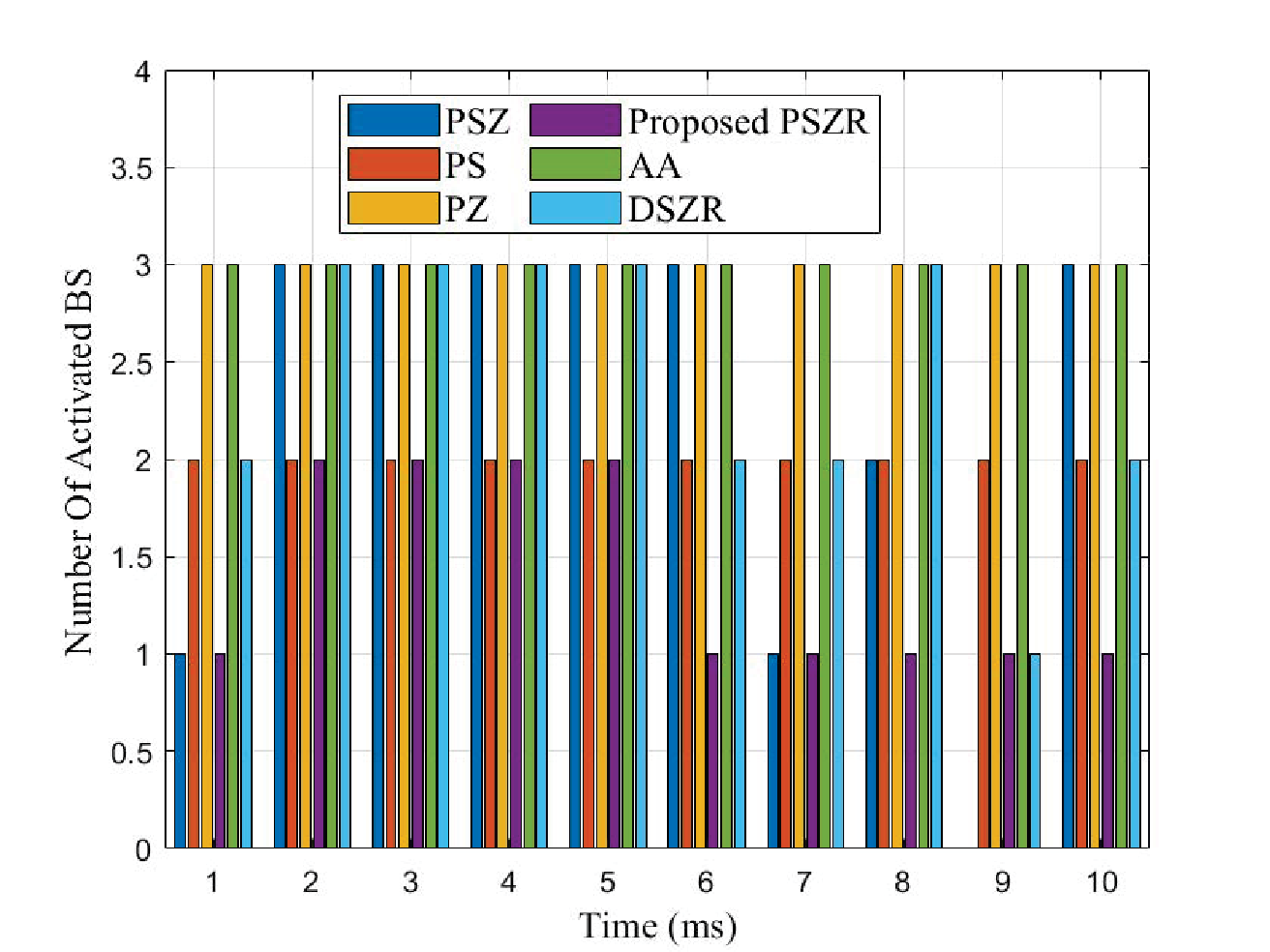}
\caption{Number of active BSs.}
\label{F4.11}
\end{figure}

Fig. \ref{F4.6} shows the comparison of energy consumption trends between the proposed scheme in this paper and the schemes with different SMs. In addition to the active state, the proposed scheme considers micro sleep (SM1), light sleep (SM2) and deep sleep (SM3). The scheme AA, in which all BSs are always active, and DSZR are used as benchmarks. The other two comparison schemes are PZR with only SM1 and PZR with both SM1 and SM2. The average traffic arrival time of the four schemes in the figure is set to 30 ms, and the traffic size is 0.02 MB. It can be seen that the cumulative energy consumption of all schemes increases over time. As the complexity of SM increases, the energy-saving effect of the scheme gradually increases. This is because as sleep depth increases, BSs have more opportunities to save energy. In addition, although the energy-saving effect of PZR with SM1 is lower than that of the baseline AA, the energy-saving effect of PZR with SM1 and SM2 is more significant. PZR with SM1 achieves about 71.07\% energy-saving compared to AA, while PZR with SM1, SM2 and the proposed scheme achieve energy-savings of approximately 79.47\% and 84.65\% respectively.

The energy-saving performance of the proposed scheme with a packet size of 0.05 MB and different average inter-arrival times (AT) from 10 ms to 35 ms is illustrated in Fig. \ref{F4.7}. It can be observed for all schemes that as the AT increases, the system's energy consumption decreases. This can be explained by the fact that as the interval time between traffic load arrivals increases, the BS can gradually enter a deeper sleep state under the optimization of the PPO algorithm. Conversely, if the interval time between traffic load arrivals is shorter, the BS can have more opportunities to enter a micro or light sleep state rather than a deeper sleep state. This demonstrates the energy-saving advantage of the proposed scheme under the low traffic load.

To further illustrate the advantages of the proposed PPO-based energy-saving method, Fig. \ref{F4.11} intuitively shows the number of active BSs over the time of different schemes under the same packet size and average inter-arrival time. First, it can be observed that at each time slot, the number of active BSs of the proposed PSZR is the smallest or one of the smallest. This can be explained by the fact that the proposed method reduces the energy consumption by minimizing the number of active BSs while ensuring the delay constraint.
\begin{figure}[t!]
\includegraphics[width=\columnwidth]{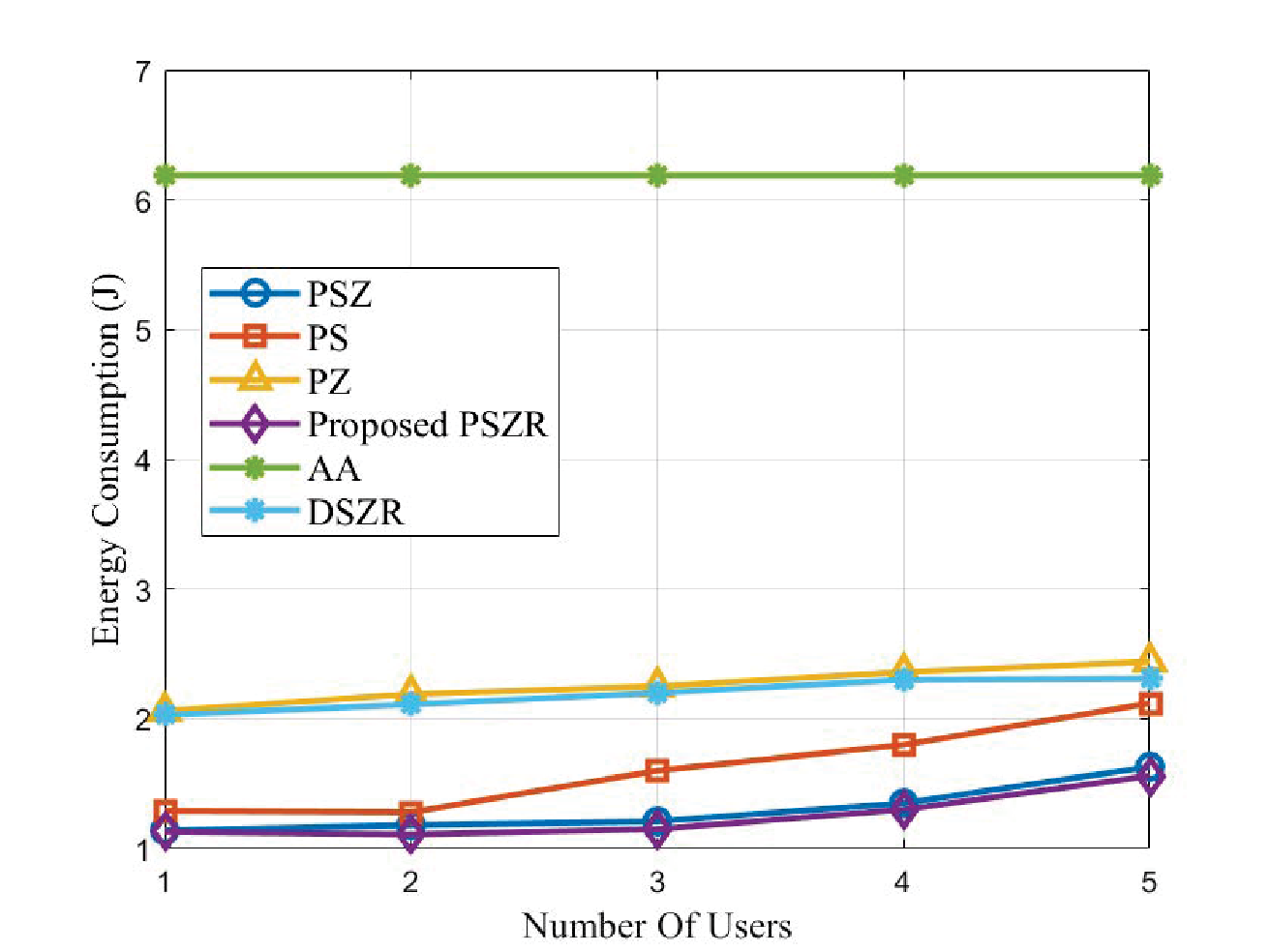}
\caption{The system energy consumption over the number of users.}
\label{usernumber}
\end{figure}

Fig \ref{usernumber} illustrates the impact of the number of users on system energy consumption. It is evident that as the number of users increases, the energy consumption of the system also increases. Compared to the benchmarks PSZ, PS, PZ, AA, and DSZR, the proposed PSZR scheme achieves the lowest energy consumption, demonstrating its superior energy-saving performance. Moreover, the energy consumption in the AA scheme remains constant because it does not adopt any energy-saving techniques or optimization strategies, keeping the BSs in an active state at all times. In contrast, PZ and DSZR show a slower increase in energy consumption as the number of users grows, indicating that these schemes are less sensitive to user count and have limited energy-saving effectiveness. The energy consumption of PS rises significantly with an increase in the number of users, reflecting that SM is user-sensitive, adjusting the BS working state based on traffic load to save energy. Compared to PS, the proposed PSZR shows a slower increase in energy consumption with more users, highlighting the additional energy-saving benefits of cell zooming, RIS, and user association.

\section{Conclusion}
\label{secconclu}
This paper proposed a novel SM method assisted by cell zooming, user association and RIS to reduce energy consumption allowing BSs to enter a sleep state during periods of low traffic load, thereby conserving energy. Cell zooming and user association enable a trade-off between energy efficiency and delay through collaborative zooming in and out among multiple BSs.  Furthermore, the RIS-assisted method further enhances energy-savings, which improves signal quality under the same transmit power by reconstructing the transmission environment. The optimization of this complex problem is achieved in two stages. First, we employ a PPO-based algorithm to optimize the SM, cell zooming, and user association. Subsequently, we proposed a new DCCN algorithm to optimize the RIS reflection coefficients. Simulation results demonstrate that our proposed approach significantly enhances the overall energy efficiency of the system while maintaining an optimal balance between energy consumption and delay.

\bibliographystyle{ieeetr}
\bibliography{ref}


\begin{IEEEbiography}[{\includegraphics[width=1in,height=1.25in,clip]{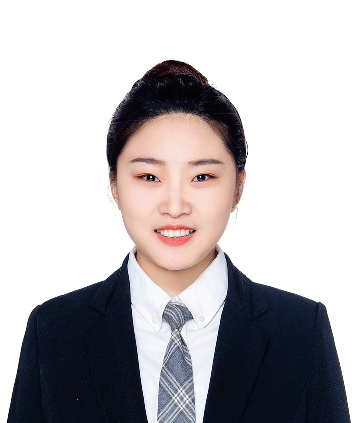}}]{Shuo Sun}
	 (Student Member, IEEE) received an M.Sc. degree in information technology from the University of New South Wales in 2022. She is currently a Ph.D. student at the Institute for Communication Systems, 5GIC \& 6GIC, University of Surrey. Her research interests are in green wireless communication.
\end{IEEEbiography}

\begin{IEEEbiography}[{\includegraphics[width=1in,height=1.25in,clip,keepaspectratio]{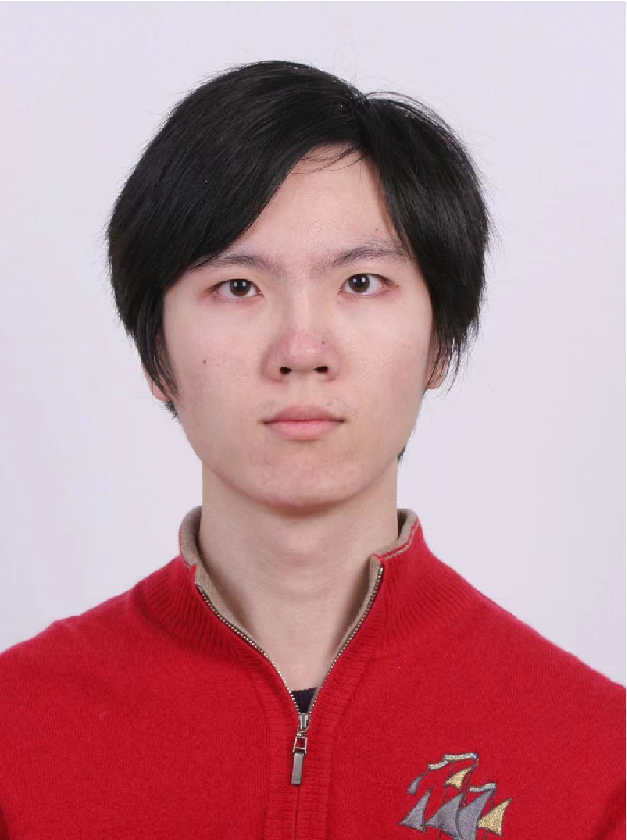}}]{Chong Huang}
(Member, IEEE) received the B.Sc. degree in communication engineering from the Nanjing University of Posts and Telecommunications in 2011, the M.Sc. degree in electrical and electronic engineering from the Loughborough University in 2015, and the Ph.D. degree in wireless communications from the University of Surrey in 2022. He is currently a research fellow with the Institute for Communication Systems (ICS), University of Surrey. His research interests include machine learning, deep reinforcement learning, cooperative networks, edge computing, satellite communications, green communications, physical layer security, cognitive radio, and reconfigurable intelligent surfaces (RIS).
\end{IEEEbiography}

\begin{IEEEbiography}[{\includegraphics[width=1in,height=1.25in,clip,keepaspectratio]{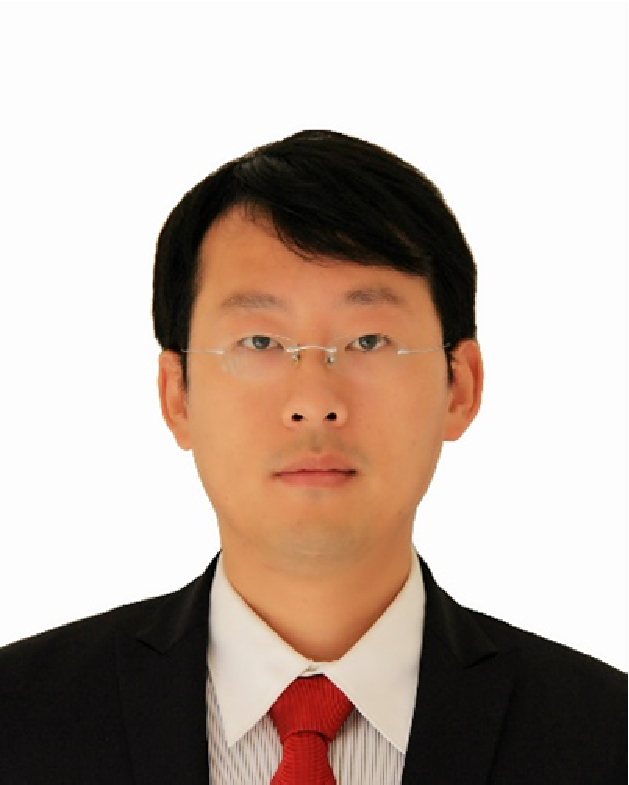}}]{Gaojie Chen}
(S'09 -- M'12 -- SM'18) received the B.Eng. and B.Ec. Degrees in electrical information engineering and international economics and trade from Northwest University, China, in 2006, and the M.Sc. (Hons.) and PhD degrees in electrical and electronic engineering from Loughborough University, Loughborough, U.K., in 2008 and 2012, respectively. After graduation, he took up academic and research positions at DT Mobile, Loughborough University, University of Surrey, University of Oxford and University of Leicester, U.K. He is a Professor and Associate Dean of the School of Flexible Electronics (SoFE), at Sun Yat-sen University, China, and a visiting professor at the 5GIC\&6GIC, University of Surrey, UK. His research interests include wireless communications, flexible electronics, satellite communications, the Internet of Things and secrecy communications. He received the Best Paper Awards from the IEEE IECON 2023, and the Exemplary Reviewer Awards of the {\scshape IEEE Wireless Communications Letters} in 2018, the {\scshape IEEE Transactions on Communications} in 2019 and the {\scshape IEEE Communications Letters} in 2020 and 2021; and Exemplary Editor Awards of the {\scshape IEEE Communications Letters}, {\scshape IEEE Wireless Communications Letters}, and {\scshape IEEE Transactions on Cognitive Communications Networking} in 2021, 2022, 2023 and 2024, respectively. He served as an Associate Editor for the {\scshape IEEE Journal on Selected Areas} in Communications - Machine Learning in Communications from 2021 to 2022. He serves as an Editor for the {\scshape IEEE Transactions on Wireless Communications}, {\scshape IEEE Transactions on Cognitive Communications Networking}, {\scshape IEEE Wireless Communications Letters}, and a Senior Editor for the {\scshape IEEE Communications Letters}, and a Panel Member of the Royal Society’s International Exchanges, UK.
\end{IEEEbiography}

\begin{IEEEbiography}[{\includegraphics[width=1in,height=1.25in,clip]{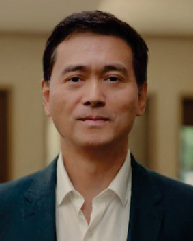}}]{Pei Xiao}
	  (Senior Member, IEEE) is a Professor in Wireless Communications in the Institute for Communication Systems (ICS) at University of Surrey. He is currently the technical manager of 5GIC/6GIC, leading the research team in the new physical layer work area, and coordinating/supervising research activities across all the work areas (\url{https://www.surrey.ac.uk/institute-communication-systems/5g-6g-innovation-centre}). Prior to this, he worked at Newcastle University and Queen’s University Belfast. He also held positions at Nokia Networks in Finland. He has published extensively in the fields of communication theory, RF and antenna design, signal processing for wireless communications, and is an inventor on over 15 recent patents addressing bottleneck problems in 5G/6G systems.

\end{IEEEbiography}

\begin{IEEEbiography}[{\includegraphics[width=1in,height=1.25 in,clip]{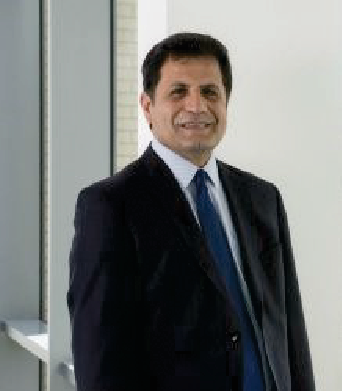}}]{Rahim Tafazolli}
	
(Fellow, IEEE) is Regius Professor of Electronic Engineering, Professor of Mobile and Satellite Communications, Founder and Director of 5GIC, 6GIC and ICS (Institute for Communication System) at the University of Surrey. He has over 30 years of experience in digital communications research and teaching. He has authored and co-authored more than 1000 research publications and is regularly invited to deliver keynote talks and distinguished lectures to international conferences and workshops. He was the leader of study on “grand challenges in IoT" (Internet of Things) in the UK, 2011-2012, for RCUK (Research Council UK) and the UK TSB (Technology Strategy Board). He is the Editor of two books on Technologies for Wireless Future (Wiley) vol. 1, in 2004 and vol. 2, in 2006. He holds Fellowship of Royal Academy of Engineering, Institute of Engineering and Technology (IET) as well as that of Wireless World Research Forum. He was also awarded the 28th KIA Laureate Award- 2015 for his contribution to communications technology.

\end{IEEEbiography}

\end{document}